\documentclass[twocolumn]{aastex61}
\usepackage{subfigure}

\shortauthors{Capparelli et al.}

\begin{document}

\title{H${\alpha}$ and H${\beta}$ emission in a C3.3 solar flare:\\
 comparison between observations and simulations}


\author{Vincenzo Capparelli}
\affil{Dipartimento di Fisica e Astronomia -- Sezione Astrofisica, Universit\'{a} di Catania,
           via S. Sofia 78, 95123 Catania, Italy}

\author{Francesca Zuccarello}
\affiliation{Dipartimento di Fisica e Astronomia -- Sezione Astrofisica, Universit\'{a} di Catania,
           via S. Sofia 78, 95123 Catania, Italy}

\author{Paolo Romano}
\affiliation{INAF -- Osservatorio Astrofisico di Catania, via S. Sofia 78, I-95123 Catania, Italy}

\author{Paulo J.~A. Sim\~{o}es}
\affiliation{SUPA,  School of Physics \& Astronomy, University of Glasgow, G12 8QQ, Scotland, UK}

\author{Lyndsay Fletcher}
\affiliation{SUPA,  School of Physics \& Astronomy, University of Glasgow, G12 8QQ, Scotland, UK}

\author{David Kuridze}
\affiliation{Astrophysics Research Centre, School of Mathematics \& Physics, Queen's University Belfast, Belfast, BT7 1NN, UK}

\author{Mihalis Mathioudakis}
\affiliation{Astrophysics Research Centre, School of Mathematics \& Physics, Queen's University Belfast, Belfast, BT7 1NN, UK}

\author{Peter H. Keys}
\affiliation{Astrophysics Research Centre, School of Mathematics \& Physics, Queen's University Belfast, Belfast, BT7 1NN, UK}

\author{Gianna Cauzzi}
\affiliation{INAF -- Osservatorio Astrofisico di Arcetri, Largo E. Fermi 5, I-50125 Firenze, Italy}

\author{Mats Carlsson}
\affiliation{Institute of Theoretical Astrophysics, University of Oslo, P.O. Box 1029 Blindern, NO-0315 Oslo, Norway}

\begin{abstract}

The Hydrogen Balmer series is a basic radiative loss channel from the flaring solar chromosphere. We report here on the analysis of an extremely rare set of simultaneous observations of a solar flare in the H${\alpha}$ and H${\beta}$ lines at high spatial and temporal resolution, which were acquired at the Dunn Solar Telescope.  Images of the C3.3 flare (SOL2014-04-22T15:22) made at various wavelengths along the H${\alpha}$ line profile by the Interferometric Bidimensional Spectrometer (IBIS)  and in the  H${\beta}$ with the Rapid Oscillations in the Solar Atmosphere (ROSA) broadband imager are analyzed to obtain the intensity evolution. The H${\alpha}$ and H${\beta}$ intensity excesses in three identified flare footpoints are well correlated in time. We examine the ratio of  H${\alpha}$ to H${\beta}$ flare excess, which was proposed by previous authors as a possible diagnostic of the level of electron beam energy input. In the stronger footpoints, the typical value of the the  H${\alpha}$/H${\beta}$ intensity ratio observed is $\sim 0.4-0.5$, in broad agreement with values obtained from a RADYN non-LTE  simulation driven by an electron beam with parameters constrained (as far as possible) by observation. The weaker footpoint has a larger H${\alpha}$/H${\beta}$ ratio, again consistent with a RADYN simulation but with a smaller energy flux. The H${\alpha}$ line profiles observed have a less prominent central reversal than is predicted by the RADYN results, but can be brought into agreement if the H${\alpha}$-emitting material has a filling factor of around 0.2--0.3.

\end{abstract}

\keywords{Sun: activity - Sun: flares - Sun: photosphere - Sun: chromosphere - Techniques: high angular resolution}

\section{Introduction} \label{sec:intro}

Solar flares are explosive phenomena occurring in the solar atmosphere, which indicate a rapid conversion of magnetic energy into other forms of energy (kinetic, radiative, particle acceleration, etc.). This process, which is believed to result from magnetic reconnection within a region with highly unstable magnetic field configurations, can produce emission of electromagnetic radiation in almost the entire electromagnetic spectrum (depending on the energy involved, which can span from 10$^{28}$ to 10$^{32}$ erg) and is associated with an increase in brightness of the corona, chromosphere and, occasionally, the  photosphere (see, e.g. \cite{fletcher} and references therein, for a review). 

Several mechanisms are involved in such a broad range of electromagnetic radiation emission at various atmospheric heights. Flares are often contextualized in the well-known CSHKP flare geometry (named for \citealt{carmi,sturr,hira} and \citealt{kopp}) which suggests that when an instability sets in, magnetic reconnection takes place (usually at the coronal level) resulting in electrons and protons being accelerated. But models dealing with coronal processes have little to say about the details of the generation of flare chromospheric emission, particularly in the lower atmosphere. Initial models of atmospheric emission lines were based on empirical models of flaring atmospheres or, assuming an electron-beam plasma heating mechanism, radiative transfer simulations \citep{canfield}. However more recently we have started to turn to radiation-hydrodynamic (RHD) flare simulations (\citealt{Abbet,allred} and \citealt{kaspa09}) which model the effect of accelerated particles traveling through the lower atmospheric layers, impulsively heating the local plasma, and causing an expansion of the chromosphere in a process termed chromospheric evaporation.

To test and constrain the electron-beam energy transport model we need to identify sensitive diagnostic radiation signatures, observations of which can be compared to the output of targeted numerical simulations. In this regard, the chromosphere presents an ideal test-bed for analyzing the release and redistribution of energy from accelerated particles in this region. In particular, observation and modeling of spectral lines emergent from different layers of the chromosphere can be used to understand how the chromosphere responds to energy input at different heights, and thereby to constrain the beam properties. But such investigations also serve an additional, exploratory, purpose by helping us to identify the best ways - e.g. choices of wavelength, temporal and spectral resolution - to get the maximum diagnostic power from flare optical observations, which can be very challenging to plan and execute.

In this paper we present flare observations at high temporal and spatial resolution in the H$\alpha$ and H$\beta$ lines, with accompanying RHD simulations. Though among the strongest spectral lines emitted by flares, simultaneous observations in these two lines are very rare. This flare is therefore of interest, as we can probe the behavior of H$\alpha$ and H$\beta$ to obtain insight into flare chromospheric excitation at their different formation heights. The H$\alpha$, H$\beta$ and H$\gamma$ lines were investigated theoretically by \citet{kaspa09} who used 1-D radiation hydrodynamics and test particle modelling to simulate the propagation, scattering and collisional energy loss of an electron beam (including direct collisional excitation of the hydrogen lines by beam particles), and calculate the emergent Balmer-line radiation. They demonstrated that Balmer line intensities are expected to be correlated on sub-second timescales, and that the intensity variations in line centres and line wings are dependent on the atmospheric heating, and the parameters of the electron beams. 

It is worthwhile to stress that simultaneous H$\alpha$ and H$\beta$ observations of a flare is very rare, therefore, the results obtained from the present investigation could provide new and important insights in the comprehension of flare emission mechanisms in the relevant wavelengths and atmospheric heights. 

In this paper, we study the flare SOL2014-04-22, that occurred close to the western limb of the Sun and compare observational data acquired from ground-based and satellite instruments with the results obtained from the Radiative hydrodynamic (RADYN) code (\citealt{carlsson97}, and \citealt{allred,AllredKowalskiCarlsson:2015}) in order to investigate the behavior of H${\alpha}$ and H${\beta}$, in response to the energy injected into the chromosphere during flaring. From observations and RADYN models of energy injection by an electron beam, we obtain and compare the absolute H${\alpha}$ and H${\beta}$ intensities, and their ratio, at high temporal resolution. We find that variations of this ratio between footpoints might be due to variations of the injected energy flux between models, suggesting that a well-calibrated H${\alpha}$/H${\beta}$ ratio, and sufficiently high temporal and spatial resolution could provide information on flare energy injection. We also examine the observed and simulated H${\alpha}$ line profile; the comparison allows us to constrain the filling factor of H${\alpha}$-emitting material. 

We believe that this study is useful for planned future observations with the next generation of large aperture solar telescopes, such as the Daniel K. Inouye Solar Telescope (formerly the Advanced Technology Solar Telescope, \cite{Keil2010})  and the European Solar Telescope \citep{Collados2010}.

This paper is organized as follows: in Sect. 2 we describe the observational data and in Sect. 3 the data analysis is reported. In Sect. 4 the results obtained from the RADYN models are reported and discussed. In Sect. 5 we draw our conclusions.

\begin{figure*}[ht]
\centering
\includegraphics[width=0.9\textwidth]{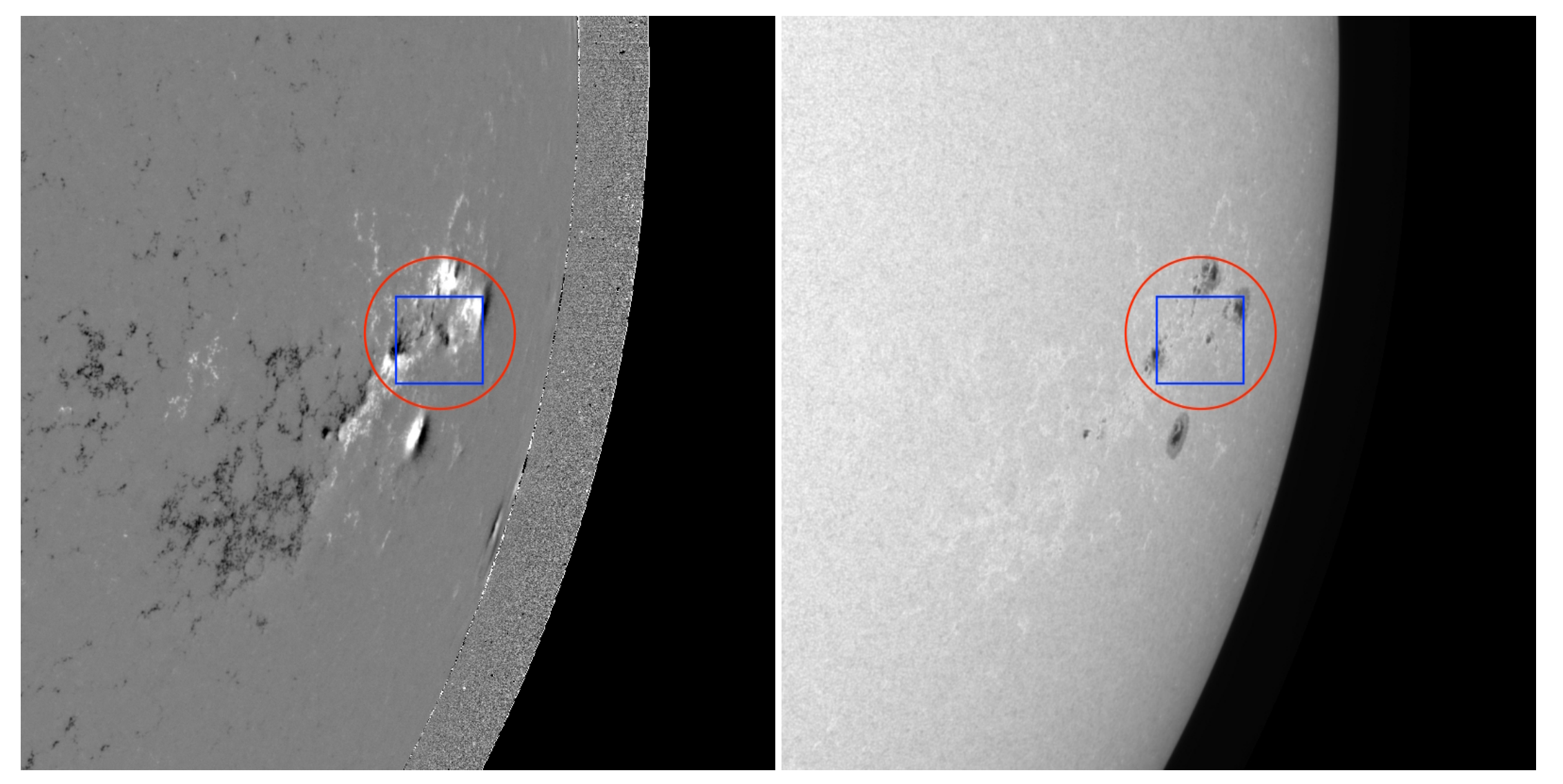} 
\caption{\textit{Left}: HMI/SDO magnetogram showing AR 12035 on 2014 Apr 22: white (black) regions indicate sites of positive (negative) longitudinal magnetic field; \textit{right}: HMI/SDO continuum image showing the photospheric configuration of AR 12035 on the same day. The red circle, with a diameter of $90''$, displays the IBIS field of view; the blue box, with a size of $ \sim 59''$, indicates the ROSA field of view. In this and in the following images, if not otherwise specified, North is on the top, West is at the right.}
\label{fig1}
\end{figure*}

\section{Observations}

An observing campaign was carried out on 2014-April-22 at the Dunn Solar Telescope (DST) at the US National Solar Observatory in New Mexico, using two different instruments: the Interferometric Bi-dimensional Spectrometer \citep[IBIS,][]{caval06} and the Rapid Oscillations in the Solar Atmosphere \citep[ROSA,][]{jess10}. The target of the observations was AR 12035, 67.7W 12.4S, characterized by a $\beta\gamma$ configuration (see Figure \ref{fig1}). 

The aim of the campaign was to determine: a) the source location, sizes and the eventual offset between flare sources at different wavelengths, including any offset between the H$\alpha$ emission in the core and in the wings, in order to investigate whether it is possible to deduce any spatial dependence of energy deposition; b) the time evolution of flare energy input by examining intensity variations on the shortest possible timescales; c) any evidence for continuum enhancement, although we don't pursue this avenue in the following.

\subsection{Ground-based observations}
The IBIS instrument acquired data in two consecutive time intervals: the first dataset was acquired during the pre-flare phase, while the second one covered all the C3.3 flare (flare B) evolution. More precisely, 
the first dataset includes 1000 scans of the H${\alpha}$ line centered at 6563 \AA~ from 14:22 UT to 15:05 UT; each spectral profile was sampled with a total of 17 wavelength points (average step = 0.2 \AA) in about 2.61 s. 
The second dataset consists of 900 scans of the H${\alpha}$ line from 15:08 UT to 15:44 UT, with the same spectral sampling used to acquire the first dataset. In both cases the H$\alpha$ 
line has been acquired in spectral mode without polarimetric measurements, with a pixel size of about $0.09''$ pixel$^{-1}$. 

For each narrow-band filtergram, a simultaneous broadband image (6610 $\pm$ 50 \AA )  was
acquired, with the same exposure time and the same field of view (FOV), characterized by a circular shape, with a diameter of $90''$.
To reduce the seeing degradation and obtain a homogeneous resolution across the whole FOV of 1000 $\times$ 1000 pixels, 
the broadband images have been restored using the Multi-Object Multi-Frame Blind Deconvolution (MOMFBD, \cite{momfb}) technique. 
We computed the global and local shifts necessary to align and destretch the broadband images with respect to the
MOMFBD restored broadband images. The same shifts have been applied to the narrow-band images. 

\begin{figure}
\centering
\includegraphics[width=0.23\textwidth]{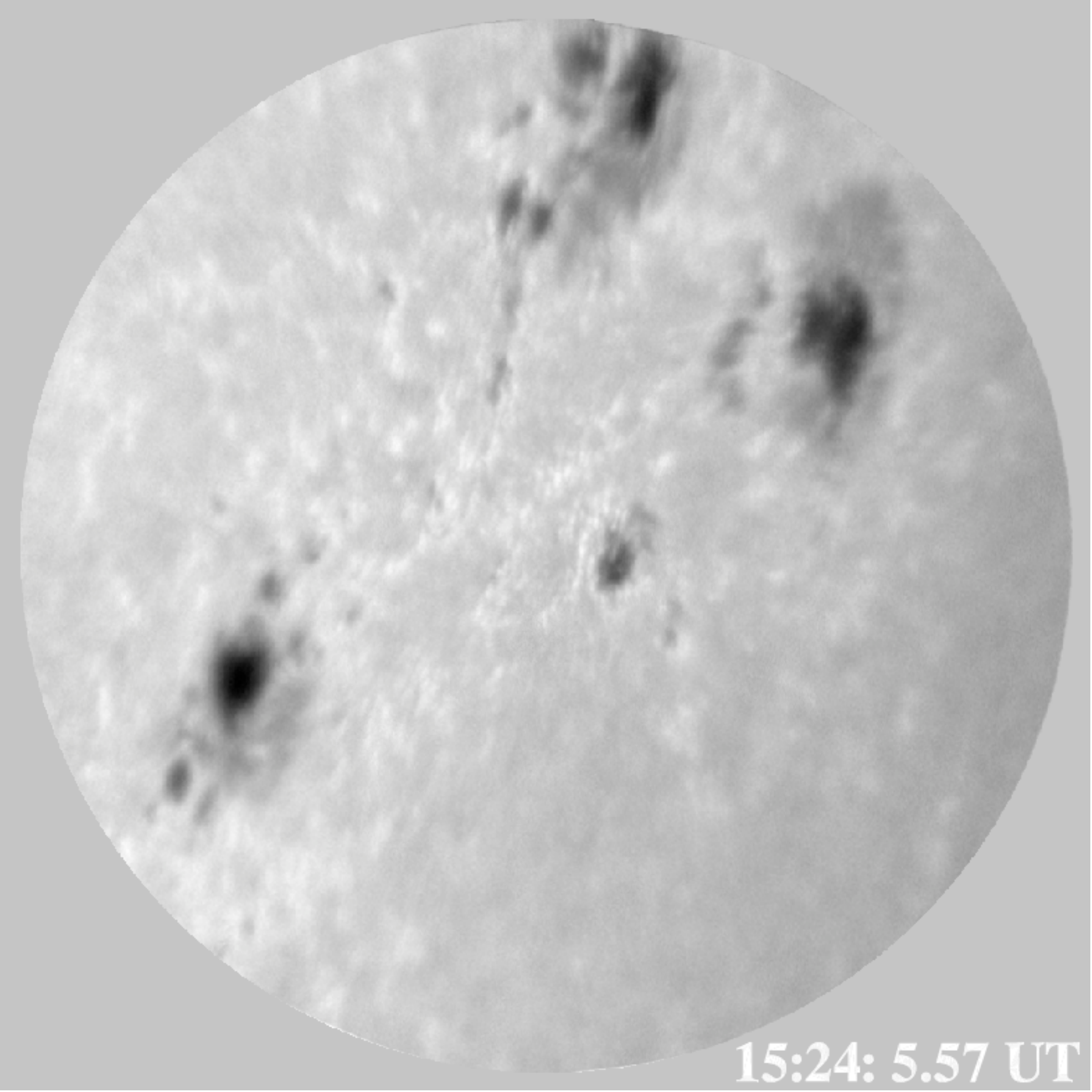}
\includegraphics[width=0.23\textwidth]{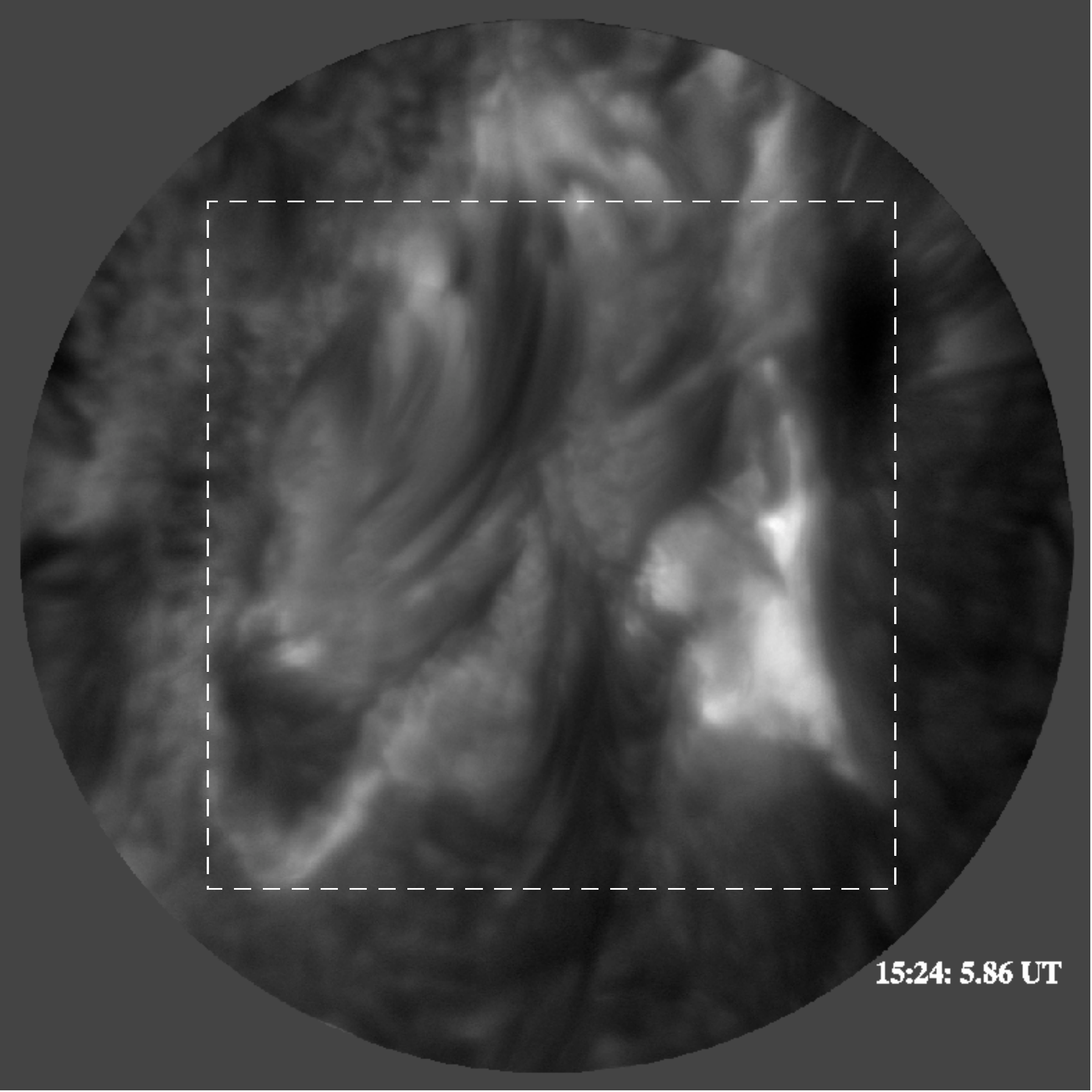}
\caption{IBIS FOV after the flare peak. ($a$) Continuum, ($b$) H$\alpha$ line core.} 
\label{ibis_con}
\end{figure}

Figure \ref{fig1} (\textit{right panel}) highlights the IBIS field of view with a red circle on SDO/HMI continuum, while Figure \ref{ibis_con} shows an example of the IBIS data in the continuum and in the core of the H${\alpha}$ line.

Co-spatial and co-temporal observations of the same active region were undertaken between 15:10:33 and 15:46:00 UT 
with the Rapid Oscillations in the Solar Atmosphere (ROSA; \citet{jess10})
imaging system installed at the DST.
The dataset includes simultaneous imaging in the Ca {\sc{ii}} K core at 3933.7 {\AA} (bandpass 1.0 {\AA}), the G band at 4305.5 {\AA}, 
(bandpass 9.2 {\AA}), continuum 4170 {\AA} (bandpass 52.0 {\AA}) and H$\beta$ core at 4861 {\AA} (bandpass 0.1 {\AA})
which was obtained through the Universal Birefringent Filter (UBF). The G band and continuum observations were obtained with a spatial sampling of $0.069''$ pixel$^{-1}$ whereas the Ca {\sc{ii}} K and H$\beta$ observations have a spatial sampling of $0.138''$ pixel$^{-1}$. The total field of view of ROSA images is $69''\times69''$, centered at S12.4 W67.3 in heliographic coordinates. High-order adaptive optics \citep{rim} were applied throughout the observations to compensate for local seeing fluctuations.

The images were reconstructed by implementing the speckle algorithms of \cite{wog} followed by de-stretching. 
These algorithms have removed the effects of atmospheric distortion from the data. The effective cadence after 
reconstruction is reduced to about 9.238 s for H$\beta$, 2.3 s for Ca {\sc{ii}} K and 2.112 s for G band and continuum.
Moreover, the FOV is reduced as a result of the Speckle reconstruction algorithm, as an apodisation windowing function is applied to the images to reduce artefacts introduced by Fourier transforms. The FOV of the reconstructed images is subsequently reduced to $58.65''$ $\times$ $58.65''$.

\begin{figure} 
\centering
\includegraphics[width=0.45\textwidth]{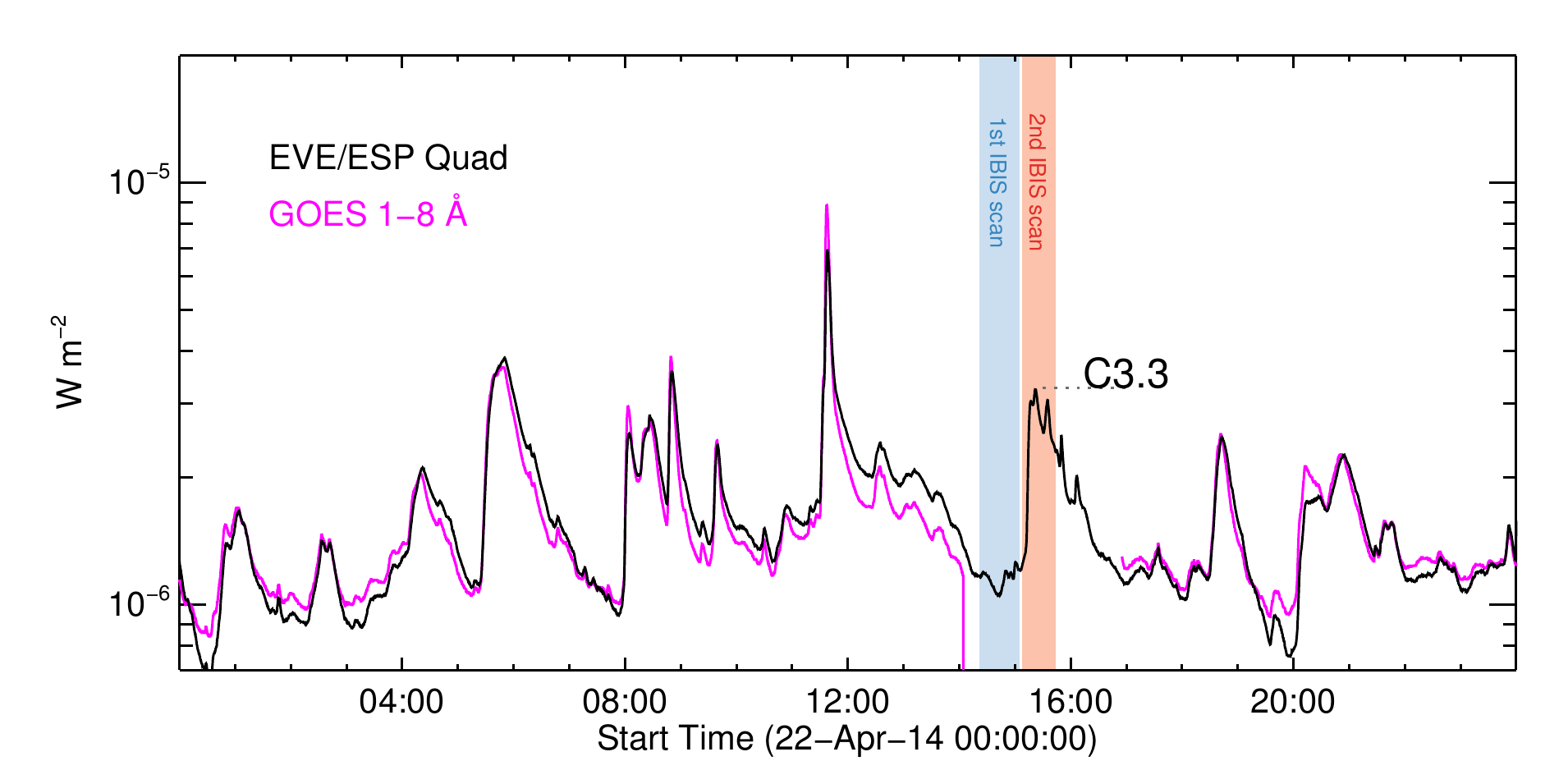}
\includegraphics[width=0.45\textwidth]{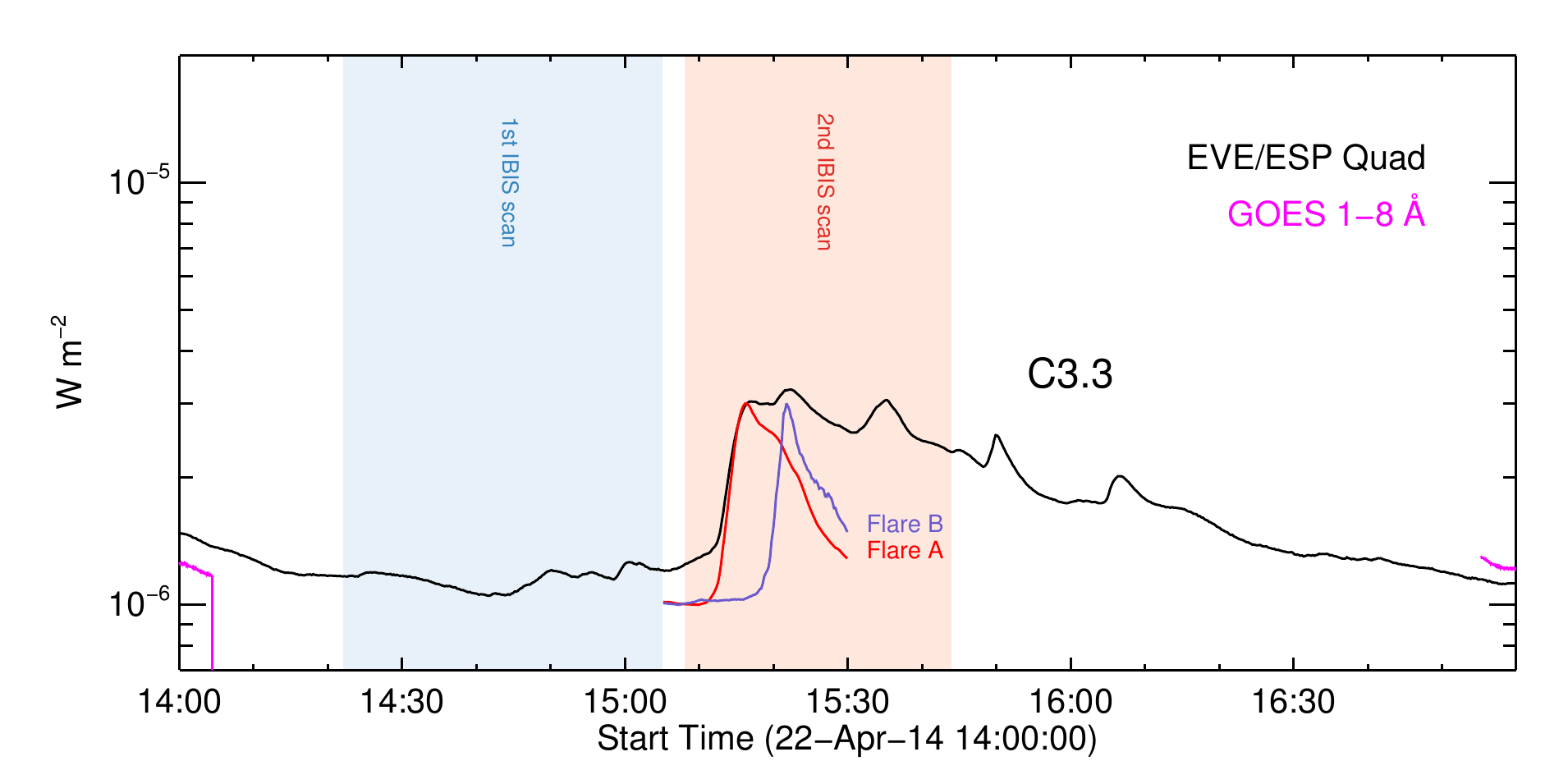}
\caption{Upper panel: Emission recorded by GOES 1-8 \AA~ and SDO/EVE ESP quad detector 1--7 \AA~ on April 22, 2014. It should be noted that between 14:10 UT and 17:00 UT GOES data are missing.  Bottom panel: enlargement of the above panel, showing the details of the emission recorded by SDO/EVE ESP. In both panels the light blue and pink bands indicate the times of IBIS acquisition. The red and violet curves show the AIA 131 \AA~ emission of flares A and B, see Figure \ref{aialight}.} 
\label{eve}
\end{figure}

\subsection{Space-based Observations}

Context images acquired with the Helioseismic and Magnetic Imager \citep[HMI,][]{Scherrer12} and the Atmospheric Imaging Assembly \citep[AIA,][]{Lemen12} instruments on-board the Solar Dynamics Observatory \citep[SDO,][]{Pesnell12} were used to provide general information on the magnetic field configuration and the morphology of the AR. Data from the quadrant diodes at 1--7 \AA~ of Euv SpectroPhotometer (ESP), part of the Extreme ultraviolet Variability Experiment \citep[EVE,][]{Woods12} on board of SDO were used to estimate the GOES classification of the flare.

In order to have information on the high energy flare emission we also used data acquired by the Reuven Ramaty High Energy Solar Spectrometer Imager (RHESSI; \cite{lin02}).

As described in the following section, a solar flare occurred in NOAA 12035, starting at 15:17 UT, peaking at 15:21 UT and ending at 15:30 UT. GOES data was not available for this event, however EVE/ESP data, represents a good proxy for GOES data and allows us to classify it as a C3.3 flare \citep{hock13}. It is important to hightlight that few minutes before, another flare occurred at the solar limb very close to NOAA 12035. To differentiate these events, we indicate with the letter B the C3.3 flare under analysis, while the letter A indicates the flare which occurred previously.  

Full-disk continuum images and longitudinal magnetograms taken by HMI on board the SDO in the Fe I line at 6173 \AA~ with a resolution of $1''$ were used to complement the high-resolution data set of the ground-based instruments.

The SDO/HMI images were aligned, taking into account the solar differential rotation, by using the IDL SolarSoft package \citep{Freeland98}.

Data taken by the Atmospheric Imaging Assembly (AIA; \cite{Lemen12}) aboard the SDO mission were used to study in detail the temporal evolution of the flare in the coronal and upper
chromospheric layers. The AIA full Sun images were processed with the usual SSW \verb!aia_prep! routines \citep{BoernerEdwardsLemen:2012,BoernerTestaWarren:2014}. EUV and UV (1600 and 1700 \AA) images have a cadence of 12 and 24 seconds, respectively. 

We reconstructed RHESSI CLEAN images using front detectors 3 to 8, for the energy ranges 6--9 and 12--25 keV\citep{HurfordSchmahlSchwartz:2002}. A sequence of 6--9 keV images with integration time of 32 seconds, stepping every 8 seconds, were constructed to obtain the light curves of flares A and B, as described in Section \ref{sect:data}.

\section{Data analysis} \label{sect:data}

\subsection{The flare evolution}
The first IBIS dataset (from 14:22 to 15:05 UT) shows many small brightenings in the wings of the H$\alpha$ line, probably Ellerman bombs occurring in a region of magnetic flux emergence \citep{ellerman,kurakowa,nindos}.
During the acquisition of the second data set (from 15:08 UT to 15:44 UT) in the south-west quadrant of the Sun two flaring regions were observed: a limb flare (flare A) in AR 12036 (start time 15:11:34 UT, end time 15:30:22 UT, peak time 15:16 UT) and a flare event (flare B) in AR 12035 (start time 15:17 UT, end time 15:30 UT, peak time 15:21 UT).

\begin{figure}[!b] 
\centering
\includegraphics[scale=.53]{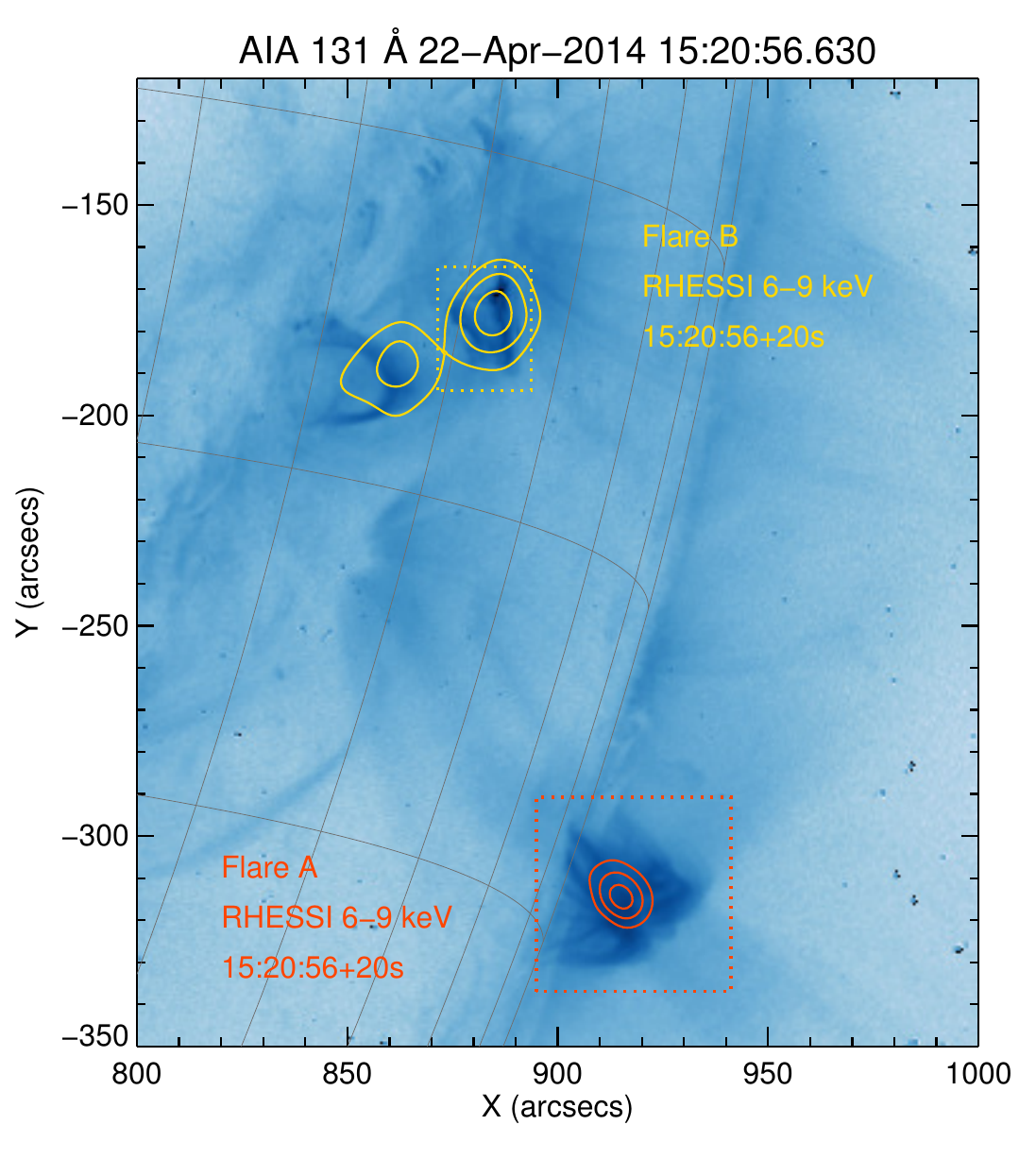} 
\caption{AIA 131 \AA~ image (reverse color) overlapped with the RHESSI contours in the 6-9 keV range, showing the location of flares A and B (see text).} 
\label{aialight}
\end{figure}

\begin{figure}[!t] 
\centering
\includegraphics[width=0.48\textwidth]{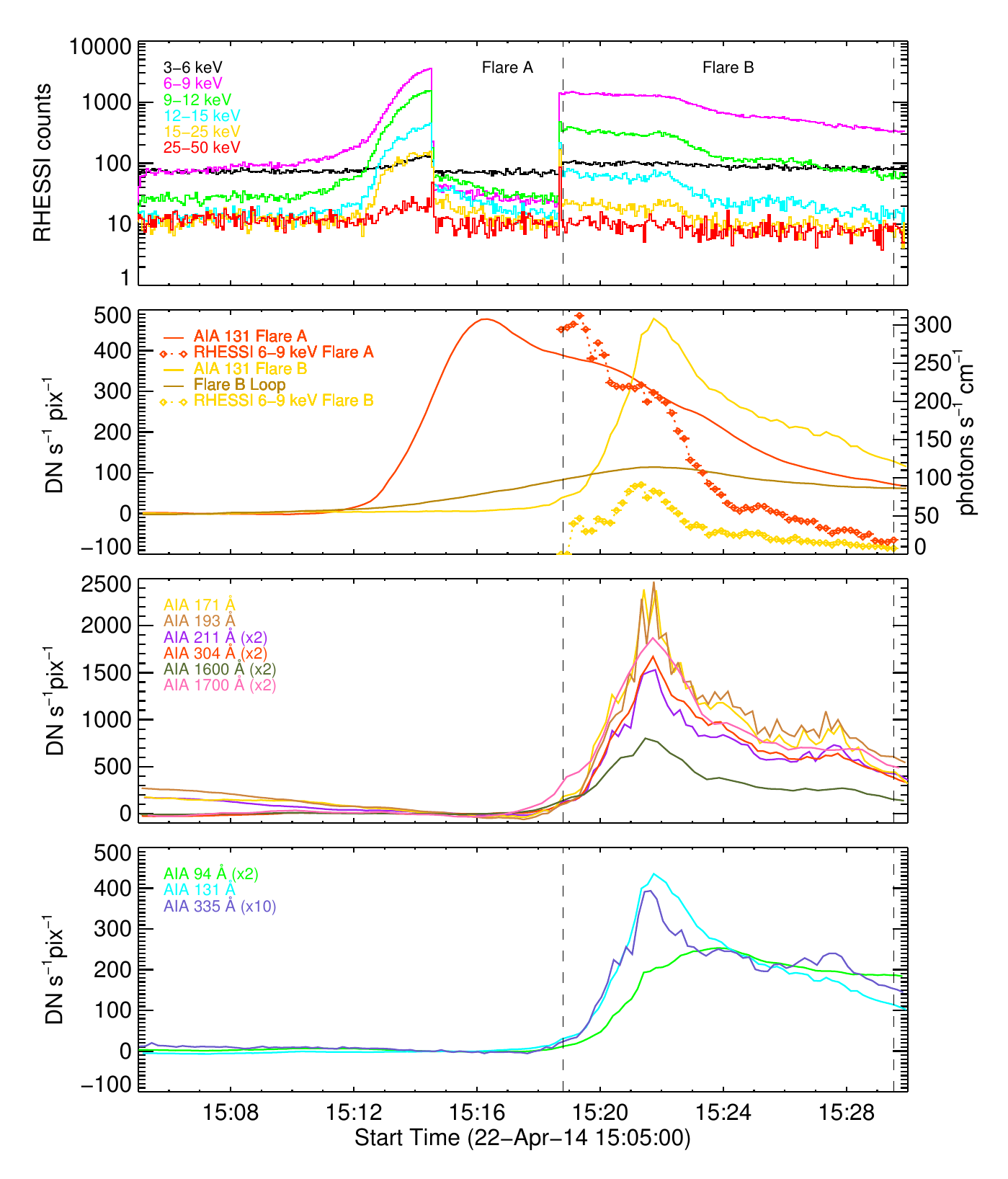} 
\caption{From top to bottom: RHESSI counts (full Sun), AIA 131 \AA~ emission of flares A and B along with RHESSI emission derived from images; AIA emission from each EUV and UV channel for flare B.} 
\label{aialight1}
\end{figure}

\begin{figure}[!b]
\centering
\includegraphics[scale=.45]{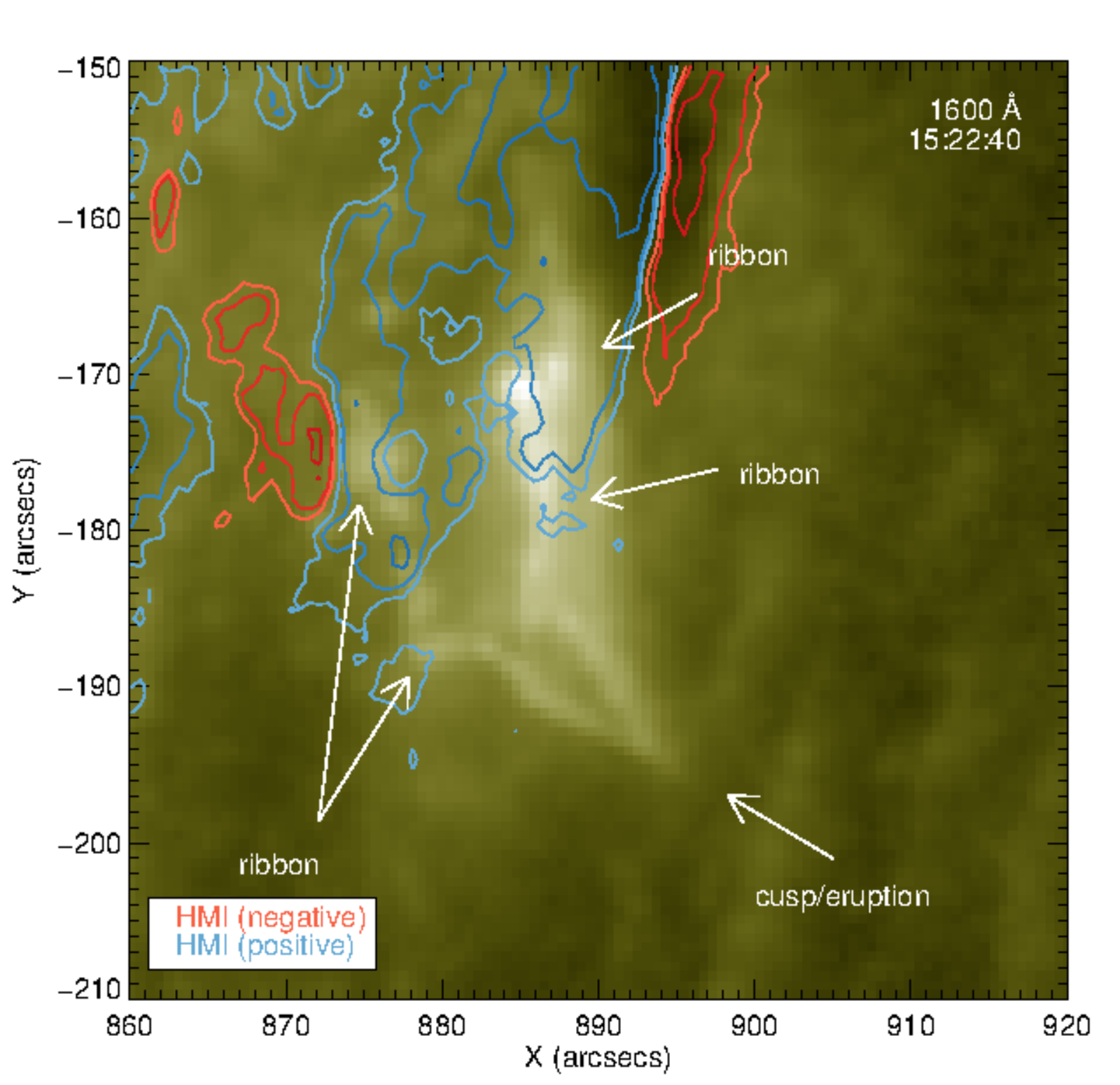}
\caption{AIA 1600 \AA~ image acquired at 15:22:40 UT overlapped with the HMI line-of-sight magnetic field (red: negative field, -2000, -1000, -500, -200, -100 G, blue: positive field, 100, 200, 500, 1000, 2000 G).}
\label{ab}
\end{figure}

\begin{figure}[!h] 
\centering
\includegraphics[scale=.5]{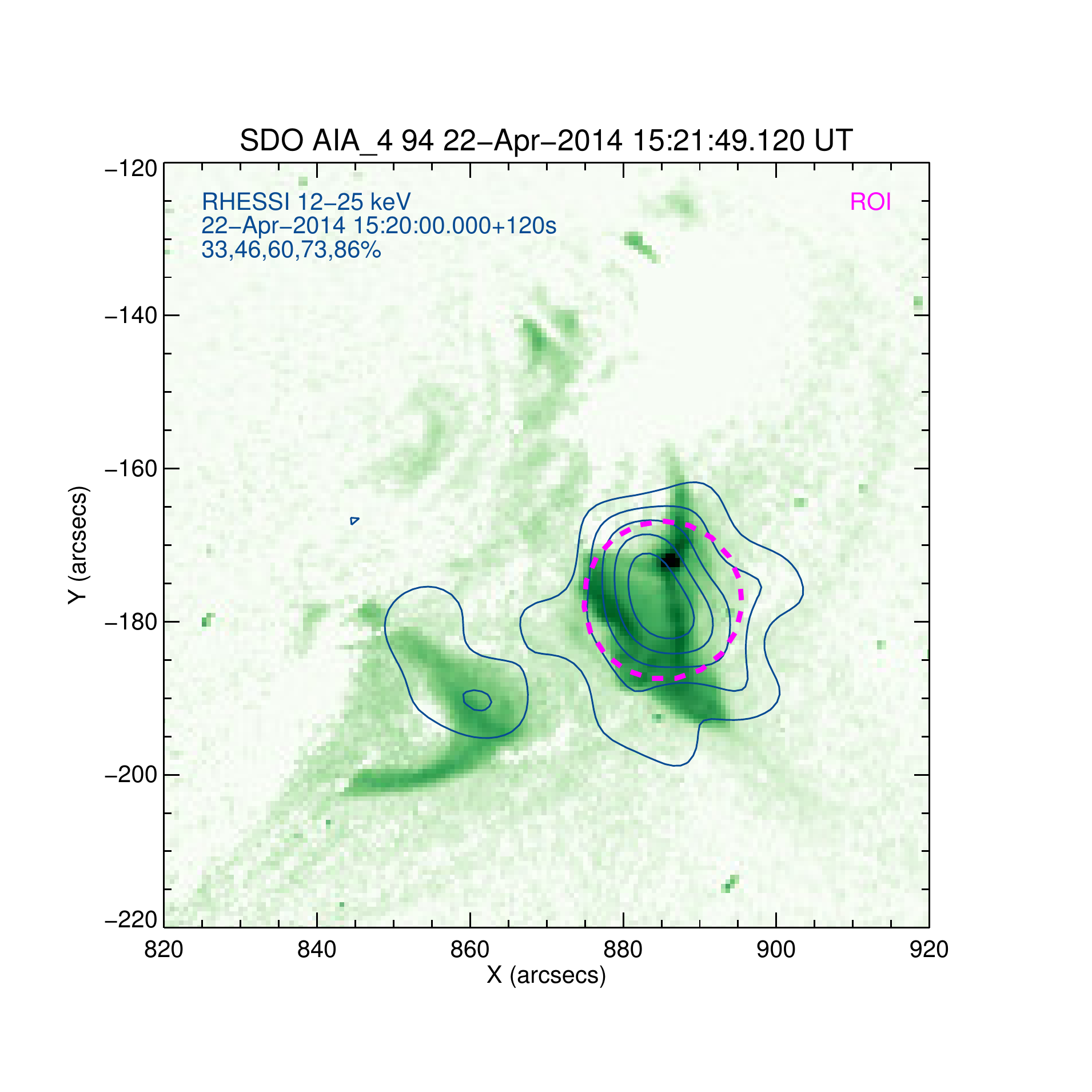} 
\caption{AIA 94 \AA~ image (reverse color) overlapped with the RHESSI emission contours in the 12--25 keV range. The purple dashed circle indicates the region of interest (ROI).} 
\label{ab1}
\end{figure}

These two flares, therefore, overlap in time, making it difficult to use full Sun data, like, for instance, the RHESSI full Sun spectral analysis or the SDO/EVE. However, using AIA and RHESSI's imaging capabilities, it is possible to infer the contributions from each flare and retrieve some information. GOES data are not available, therefore, we used the EVE/ESP quadrant diode 1--7 \AA~ as proxy (see Figure \ref{eve}).

We used the AIA 131 \AA~ channel as a reference to distinguish the two flares, because this AIA channel has a temperature response close to the plasma temperatures that can be observed by RHESSI at low energies. We were able to image each flare independently with RHESSI, as shown in Figure \ref{aialight}. To evaluate how each flare evolves, we defined a box around the location of each flare (dotted-boxes in Figure \ref{aialight}) and obtained the average emission in the AIA 131 \AA~ channel (in DN s$^{-1}$ pixel$^{-1}$).

In Figure \ref{aialight1} we show, from top to bottom: RHESSI counts (full Sun), AIA 131 \AA~ emission of flares A and B along with RHESSI emission at 6-9 keV derived from the images; AIA emission from each EUV and UV channel for flare B. The peak in the AIA 131 \AA~ channel occurs at 15:16:20 UT and 15:21:44 UT for flares A and B, respectively. From the AIA 131 \AA~ emission in the second panel of Figure \ref{aialight1} it is clear that the gradual phase of flare A extends during the impulsive phase of flare B. Therefore, we cannot use RHESSI full Sun spectroscopic analysis to study flare B.

The analysis of the IBIS and ROSA dataset allows us to follow the flare evolution in the chromosphere, which is mainly characterized by a loop-like structure with a cusp at its top (see, e.g., the right panel of Figure \ref{ibis_con}) oriented towards the south-west. The distance between the two footpoints, as determined from the IBIS H$\alpha$ image acquired at 15:24 UT (see Figure \ref{ibis_con} \textit{right} panel), is 11500 km and, assuming a semi-circular shape for the loop-like structure, this has a length of $\sim$ 18000 km. 

\begin{figure*}
\centering 
\subfigure{\includegraphics[trim=80 90 90 90, clip,width=0.28\textwidth]{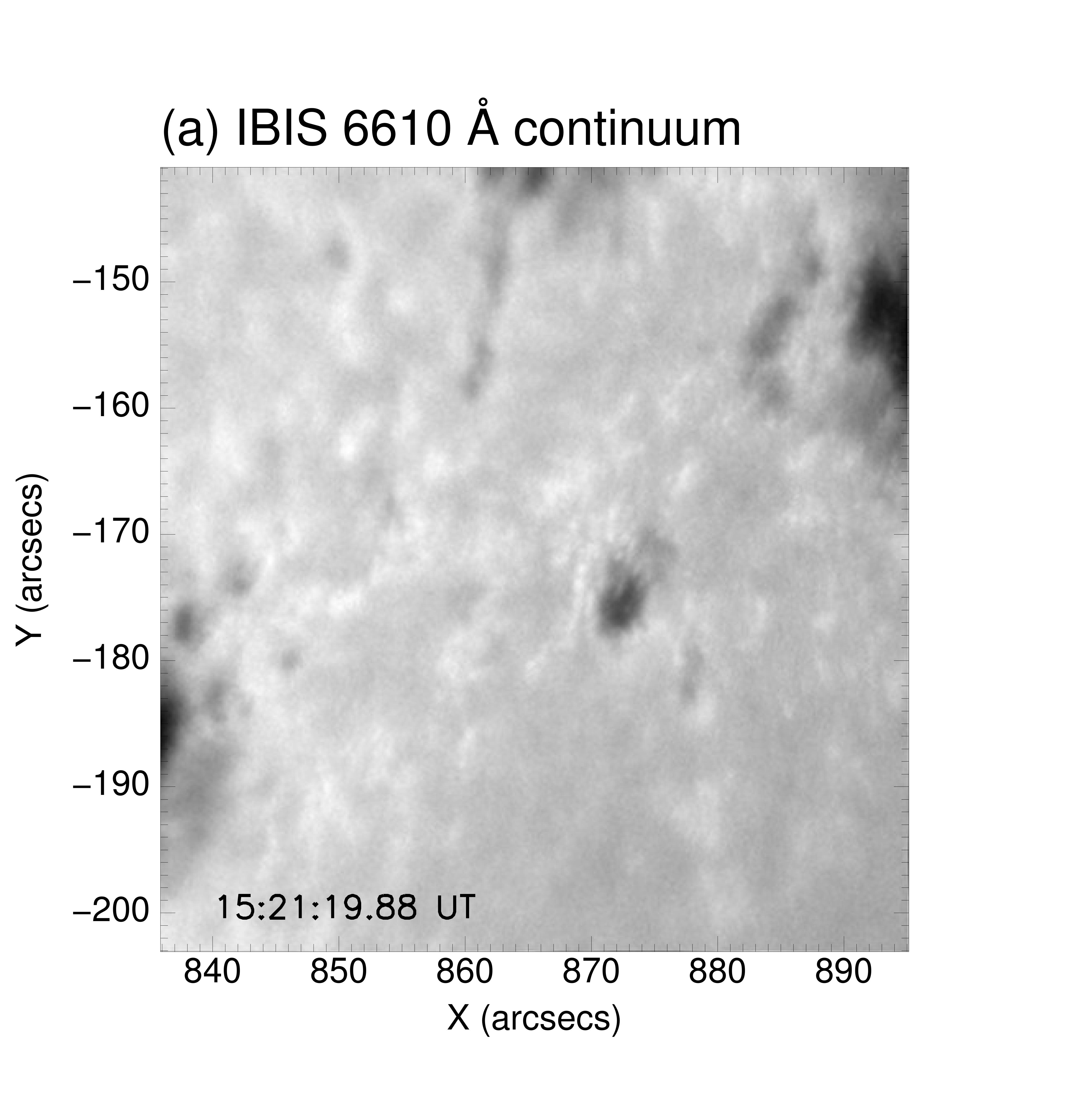}}   
\subfigure{\includegraphics[trim=80 90 90 90, clip,width=0.28\textwidth]{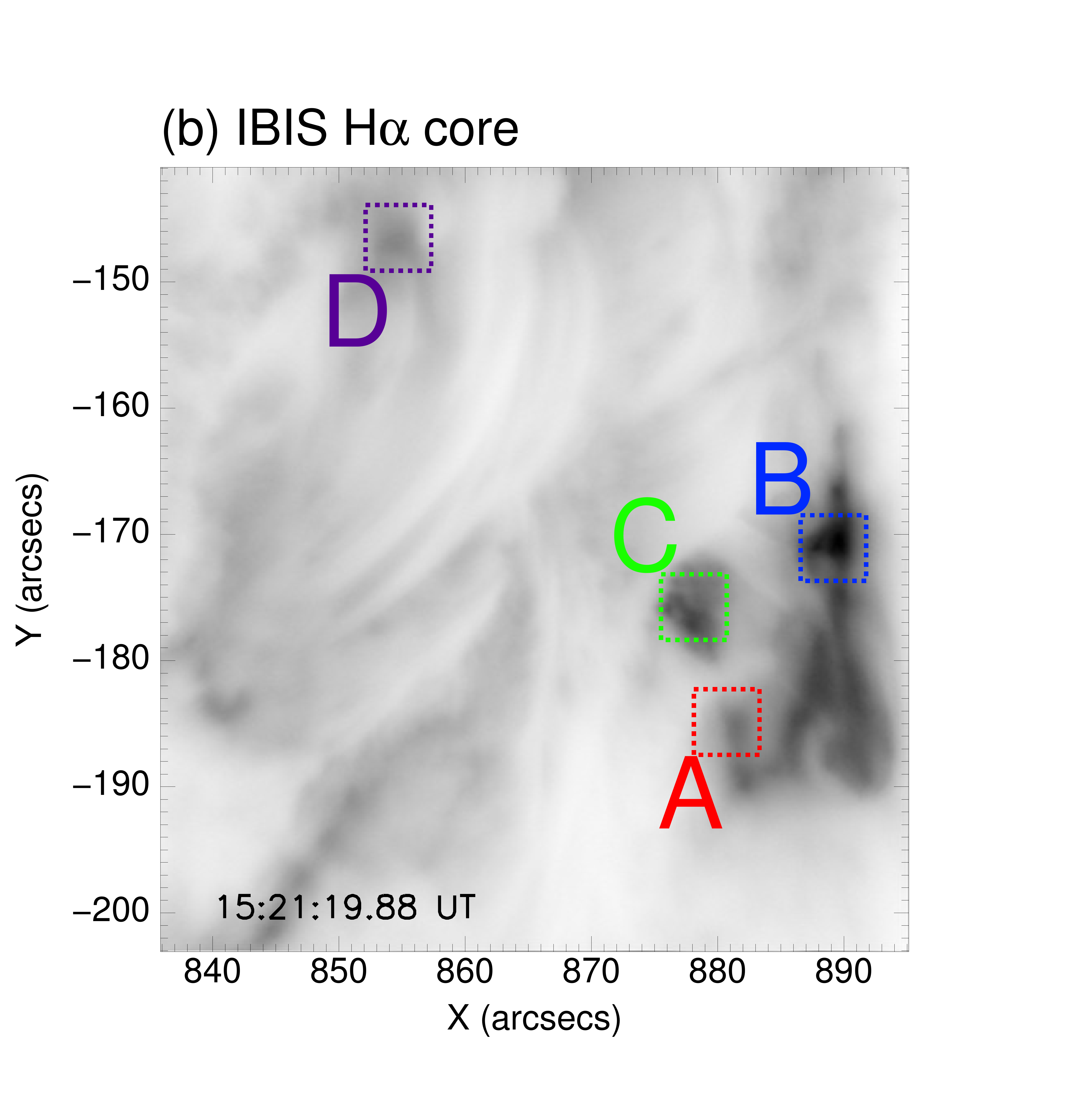}}
\subfigure{\includegraphics[trim=80 110 110 110, clip,width=0.28\textwidth]{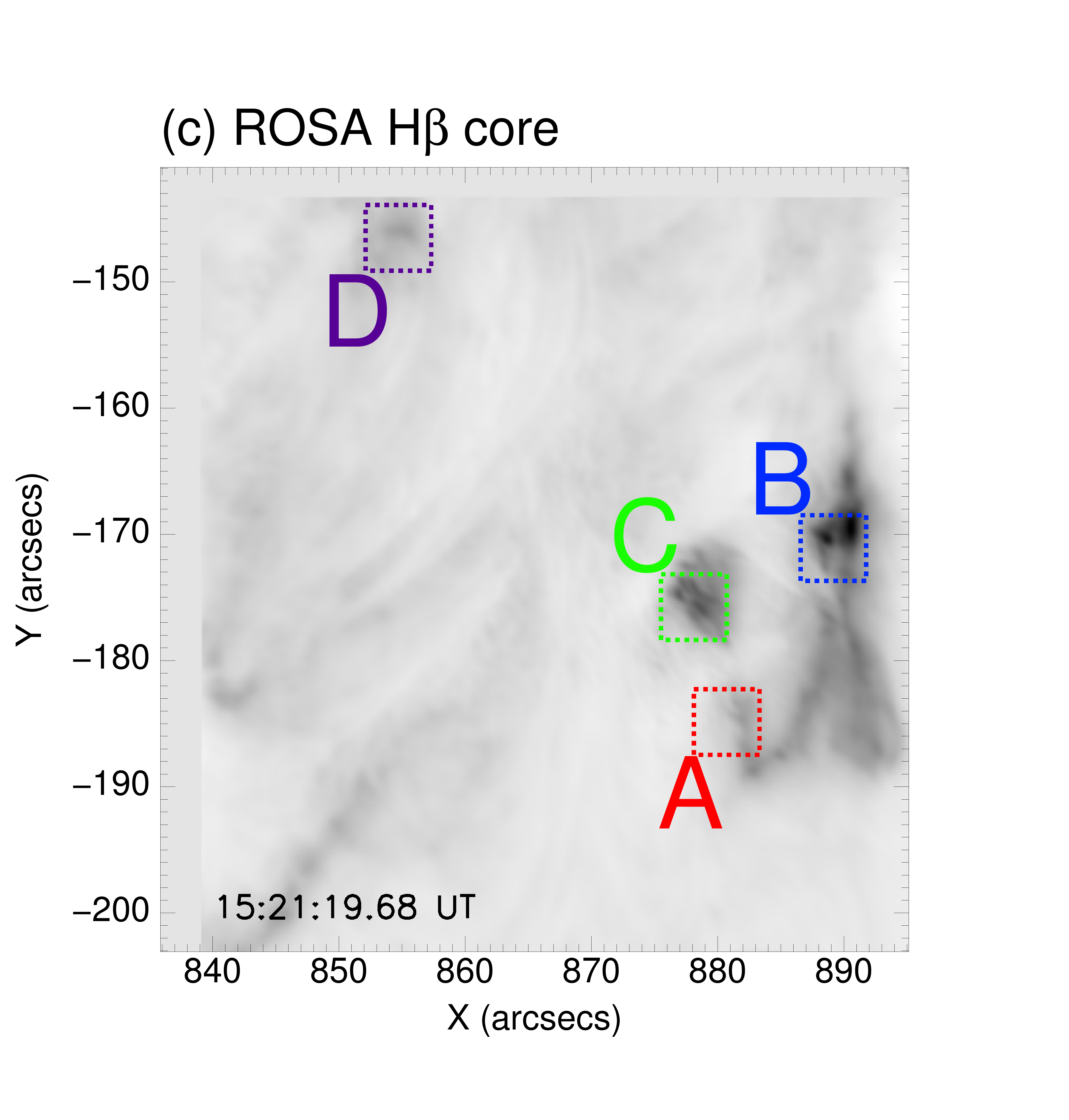}} 
\caption{(a) Image acquired in the IBIS continuum at 6610 \AA; (b) IBIS H${\alpha}$ image of the same FOV (reversed color); (c) ROSA H${\beta}$ image (inverted color) after the alignment procedure with the IBIS corresponding image. The boxes in (b) and (c) indicate the regions A, B, C and D that are used to determine the intensity evolution (see Figs. \ref{abcd} and \ref{ratio}).}
\label{align}
\end{figure*}

Comparing with continuum images (see Figure \ref{ibis_con}, \textit{left} panel), it is possible to establish that the flare developed between the large (western) sunspot and a small pore located at the center of the FOV. The large arch filament system (AFS) connecting the other two main sunspots does not seem to show any variation during the flare.

In Figure \ref{ab} we show an AIA 1600 \AA~ image acquired at the peak of the flare, with the overlapped contours (levels indicated in the figure caption) denoting the longitudinal magnetic field deduced from the HMI instrument. We notice that the cusp region, also observed at 6563 \AA~ (compare with Figure \ref{ibis_con}, \textit{right} panel), is quite evident, while the bright footpoints seem to be both located in regions of positive magnetic field, which is due to projection effects due to the proximity of the AR to the western limb.  

The comparison between the AIA 94 \AA~ image acquired $\sim 1$ minute before the flare peak (see Figure \ref{ab1}) and the RHESSI 12--25 keV contours indicate that there are two main sources of hard X-ray emission: the location of one corresponds to the flaring loop-like structure, the other is superimposed on another bright structure to the east of the flaring loop. It is worth noting that this feature was not in the IBIS and ROSA FOVs.

\subsection{H${\alpha}$ - H${\beta}$ comparison}

We further investigate the flare event through the comparison between H${\alpha}$ and H${\beta}$ images acquired by IBIS and ROSA, respectively. In fact, although this event is not very energetic and it is very close to the limb, the combined dataset is unique because it has both H${\alpha}$ and H${\beta}$ imaging.

Due to the fact that the images obtained by the two instruments have different size and spatial resolution, we first need to align the H${\alpha}$ and H${\beta}$ images. To this aim, we use the USAF target images and dot grid images (see, e.g., \citealt{kleint}), because the sample is the same during the observing campaign and, furthermore, they were simultaneously acquired in the two channels.
Through them it was possible to calculate the parameters to rotate, rescale and shift 
the images in order to obtain the correct alignment between H${\alpha}$ and H${\beta}$ images. Once obtained these parameters they have been applied 
to the IBIS dataset, i.e., the H${\alpha}$ images. Figure \ref{align} shows the result of the alignment procedure: the images have the same orientation, size (456 $\times$ 478 pixels) and 
spatial sampling ($0.138''$ pixel$^{-1}$). 

After the alignment we compared the lightcurves of these two chromospheric lines in four boxes inside the field of view indicated in Figure \ref{align} ((b) and (c) panels). Three of these boxes (A, B and C) are 
on the bright areas of the flare, while the last one (box D) is on a bright patch outside the flaring region. The boxes have the same size, i.e., 40 $\times$ 40 pixels ($\approx$ $5.5''$ $\times$ $5.5''$). We choose a box of this size to avoid to lose information during the flare brightness evolution. We note that the locations of the flare emission inside the boxes are not fixed in space but they moved during the evolution of the event. We obtained the lightcurves of H${\alpha}$ and H${\beta}$ by taking the average intensity values calculated over all pixels inside each box. Moreover, to better compare the intensities acquired with the IBIS and ROSA instruments, we calibrated the intensity obtained by the different cameras from DN to erg s$^{-1}$ cm$^{-2}$ \AA$^{-1}$ sr$^{-1}$ units as follows: we determined the flat-field image intensity acquired on the day of observation with both IBIS and ROSA cameras and, assuming that this flat-field intensity was equal to the values provided in the Brault \& Neckel Atlas \citep{brault}, we converted the measured intensity in DN to fundamental units. We reiterate here that the cadences for the reduced H${\alpha}$ and H${\beta}$ datasets are around 2.6 s and 9.2 s, respectively. 

\begin{figure}
\centering
\includegraphics[width=0.47\textwidth]{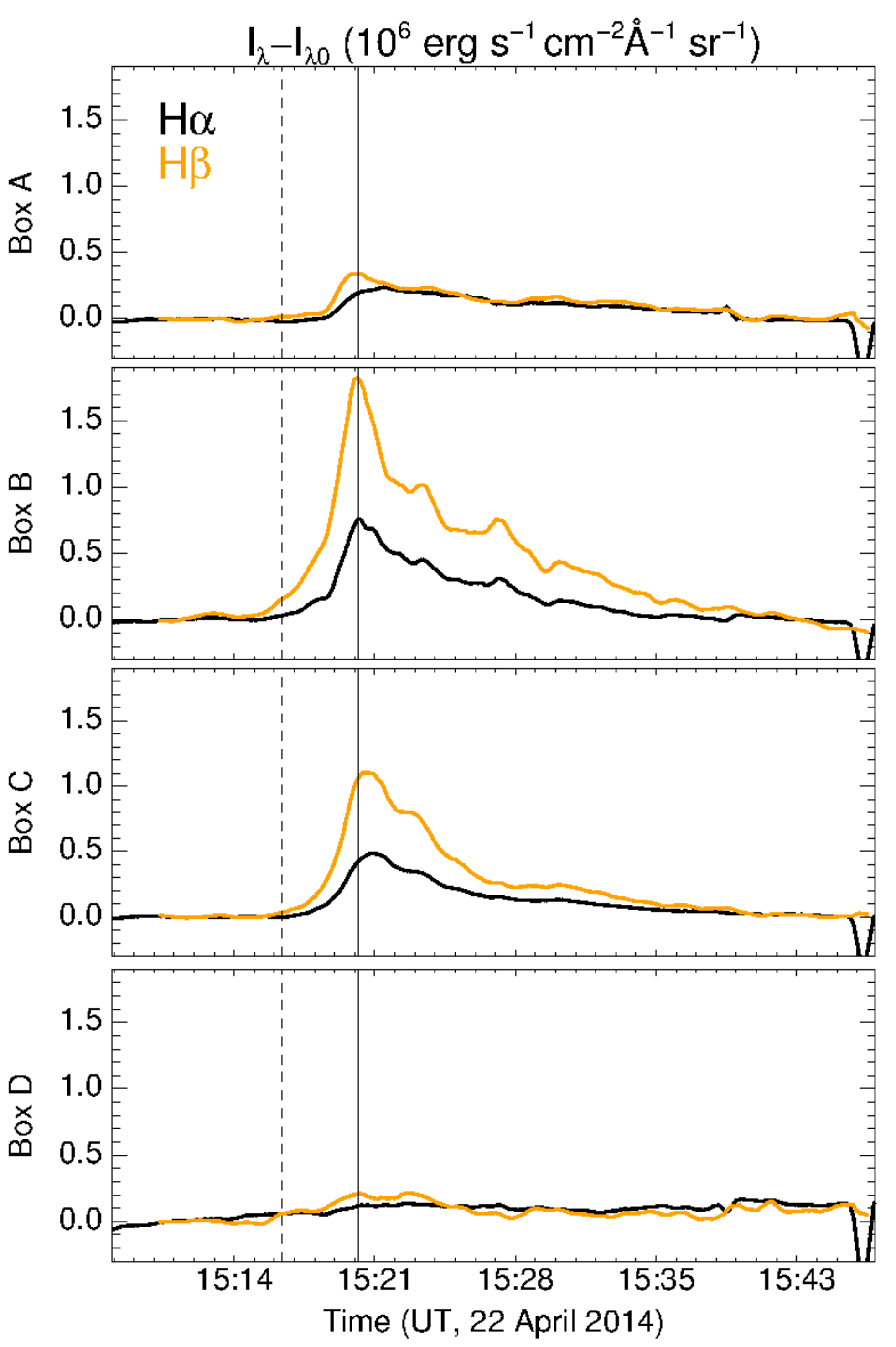}
\caption{Average intensity (after subtraction of the pre-flare intensity, see text) as a function of time deduced from H$\alpha$ (black) and H$\beta$ (orange) data in the three boxes located into the flare location (A, B and C) and in a box outside the flaring region (D). The vertical black lines show the estimated start (dashed) and peak (solid) times of the flare. The co-temporal drops in the end of sequence are related to the bad seeing conditions.}
\label{abcd}
\end{figure}

Figure \ref{abcd} shows the intensity as a function of time deduced from H$\alpha$ (black) and H$\beta$ (orange) images for the four boxes. In particular, the intensity relative to the H$\alpha$ line, has been determined by averaging the intensity at the center of the line (6562.8 \AA) and in an adjacent point along the line profile (6562.9 \AA) covering a total $\Delta \lambda = 0.1$ \AA, comparable to the bandwidth of the H$\beta$ filter employed with the ROSA instrument. However, because the cadence of H${\alpha}$ images was higher than H${\beta}$ data, the sampling to determine the H${\alpha}$ intensity was greater. To highlight the energy released during the flare, the intensity of all the lightcurves were obtained by subtracting the pre-flare intensity calculated by averaging the intensity in each box during 2 minutes in a time interval before the flare, i.e., from 15:10:00 UT to 15:12:00 UT. 

The analysis of these plots indicates that the intensity excess of the H$\beta$ line is generally higher than the H$\alpha$ intensity during the impulsive phase of the flare, in particular the former can reach values up to $\sim 0.3 \times 10^{6}$ (box A), $\sim 1.8 \times 10^{6}$ (box B) and $\sim 1.1 \times 10^{6}$ (box C) erg s$^{-1}$ cm$^{-2}$ \AA$^{-1}$ sr$^{-1}$, while the latter reaches value of $\sim 0.2 \times 10^{6}$ (box A),  $\sim 0.7 \times 10^{6}$ (box B) and $\sim 0.5 \times 10^{6}$ (box C) erg s$^{-1}$ cm$^{-2}$ \AA$^{-1}$ sr$^{-1}$ (see the first three panels of Figure \ref{abcd}).
The lightcurves obtained in the box D appear flat for both the lines during all time of analysis (see the bottom panel of Figure \ref{abcd}). In the same plot the vertical black lines show the estimated start (dashed) and peak (solid) flare times deduced by EVE/ESP.

\begin{figure}
\centering
\includegraphics[width=0.47\textwidth]{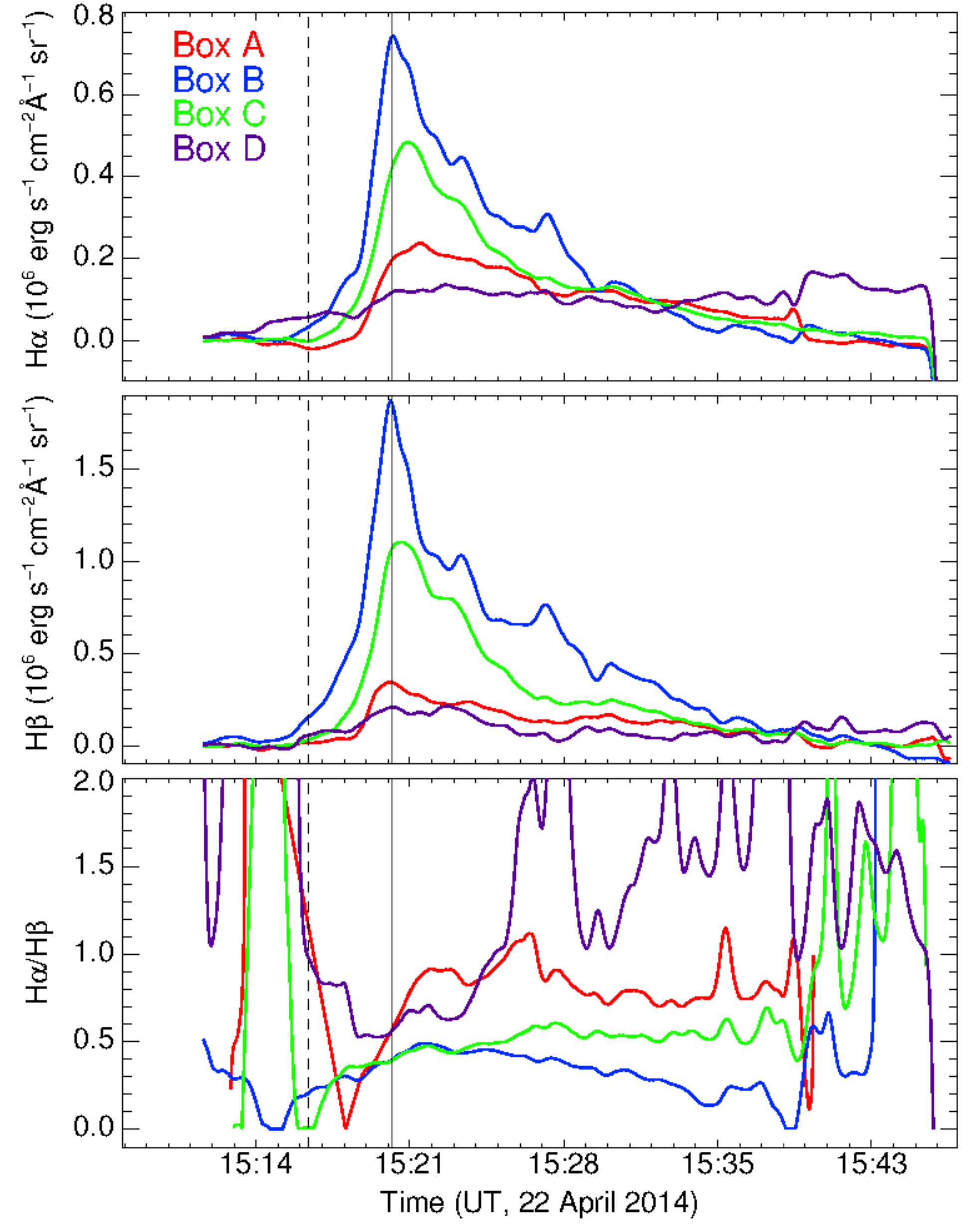}
\caption{Temporal evolution of the H$\alpha$ excess intensity (top panel) and the H$\beta$ excess intensity (middle panel) in boxes A, B, C and D indicated in Figure \ref{align} (b)-(c)). The bottom panel shows the temporal evolution of the ratio H$\alpha$/H$\beta$. Different colours indicate different boxes (see the legend in the plot). The vertical black lines show the estimated start (dashed) and peak (solid) flare times.} 
\label{ratio}
\end{figure}

The ratio of the two core spectra intensities has potential diagnostic importance for the comprehension of chromospheric flares \citep{kaspa09}. In order to detect any possible signature of different emission among the two wavelengths, we calculated the ratio between the H$\alpha$ and the H$\beta$ intensity for all the boxes in which we applied an 11-point smoothing function in order to remove the excess noise. The values of ratio have been calculated on the common acquisition time of the two instruments, namely from 15:10 UT to 15:45 UT. The bottom panel in Figure \ref{ratio} presents the temporal evolution of this ratio, where the different colours indicate the intensity ratio obtained for each box (see legend). In the same figure the intensity of the H$\alpha$ (top panel) and H$\beta$ (middle panel) lines as a function of time for each box is reported. Figure \ref{ratio} shows that the two chromospheric lines respond to the flare energy input in the same way, highlighting a similar shift between peaks in intensity and an energy distribution comparable with the corresponding box. Specifically, the H$\alpha$ lightcurves of boxes A, B and C have similar behaviour, with a similar peak in intensity and similar decay phases. Similar behaviour, albeit with different values of intensity is seen in the corresponding H$\beta$ lightcurves.

The bottom panel in Figure \ref{ratio} shows that the H$\alpha$/H$\beta$ intensity ratios during the flare tends to a constant value for the boxes inside the flaring region (approximately around 0.4 for the boxes B and C and around 0.8 for the box A), while before and after the flare the values are generally higher, with a variable trend for all the boxes. Outside the flaring region (box D) the ratio is highly unstable with large oscillations before, during and after the energy input, due to the low values of the intensity that cause large errors.

\subsection{Spatial offset}

\begin{figure}
\centering
\includegraphics[width=0.4\textwidth]{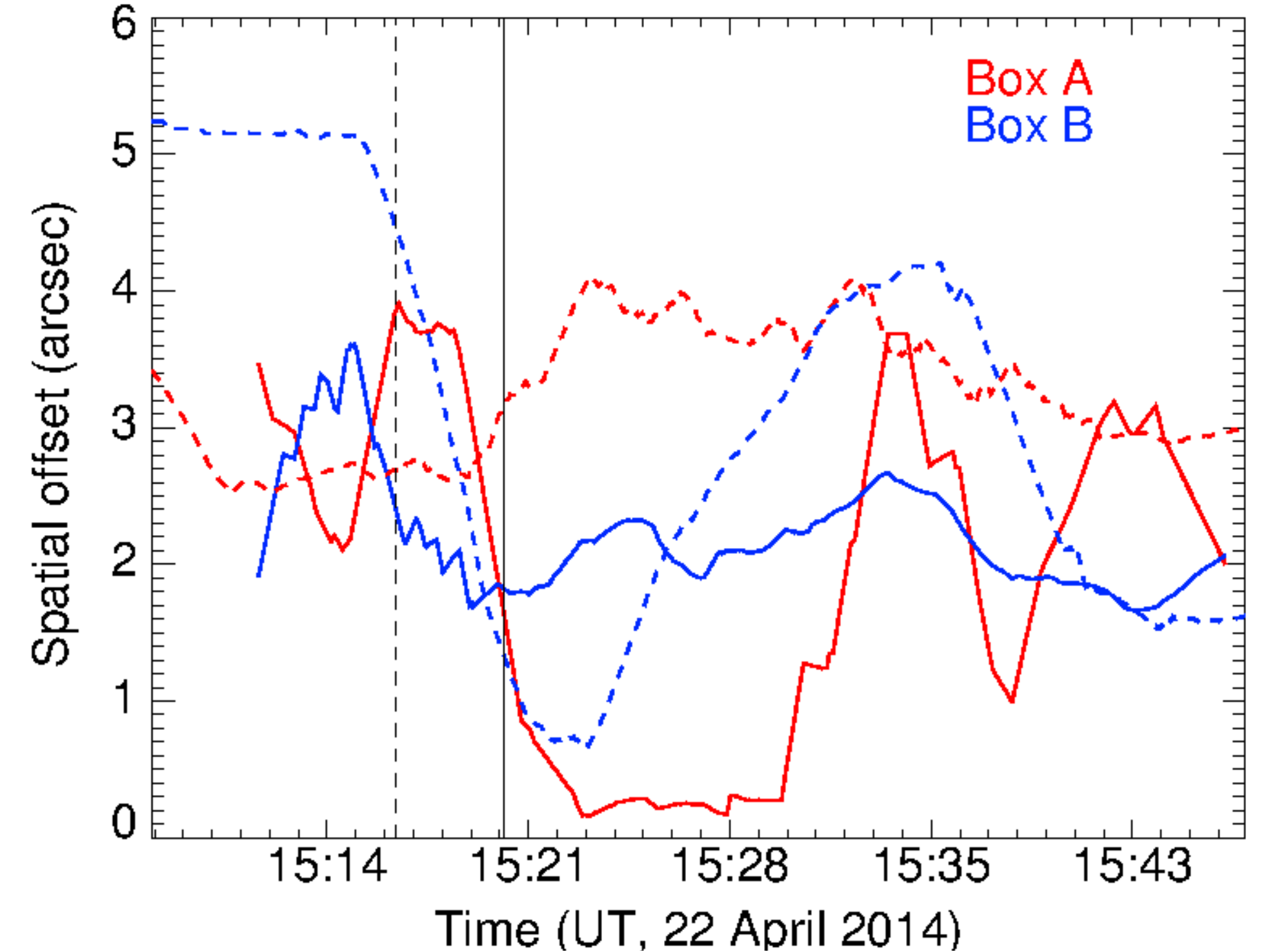}
\caption{Temporal evolution of the offset of the brightest points imaged in the H$\alpha$ and H$\beta$ line cores inside boxes A, B (solid lines). The dashed lines indicate the same, but for H$\alpha$ core line and H$\alpha$ wing.} 
\label{distance}
\end{figure}

Following the aim of the observing campaign, to evidence a possible spatial offset among each chromospheric sources, we looked at the maximum intensity value inside the corresponding box (note, it is possible that this location will not occur exactly in the same pixels for both H$\alpha$ and H$\beta$ channels). In Figure \ref{distance} we report the spatial offset as a function of time between the brightest points in the H$\alpha$ and H$\beta$ line cores for boxes A and B and the offset between the brightest points in the H$\alpha$ core and H$\alpha$ continuum wing. We display the results only for boxes A and B because they are more relevant as they are located in the footpoints of the flare loop. The analysis of this plot shows a spatial offset in the range of $2''$--$3''$ between the sources imaged in the H$\alpha$ and H$\beta$ line cores, but in the box A this offset decreases to $0.2''$ after the impulsive phase of the flare, while remains constant for the Box B. The distance between the brightest points observed in the H$\alpha$ core and H$\alpha$ wing is constant for box A around a value of $3''$, while it varies in a range of $1''$--$5''$ for box B.

\subsection{Evolution of H$\alpha$ line profile}

In order to investigate the temporal evolution of the H$\alpha$ line profile and to compare it with the simulations shown in Section 4, we selected a time interval of 10 minutes during the flare evolution, starting at 15:15:50 UT and calculated the average intensity in each point of the line acquired by IBIS (it should be noted that a similar analysis for the ROSA H$\beta$ dataset could not be performed because in this case we only have images in the line center). Again, this analysis was only carried out for boxes A and B.

\begin{figure*}[htb]
\centering
\includegraphics[width=0.90\textwidth]{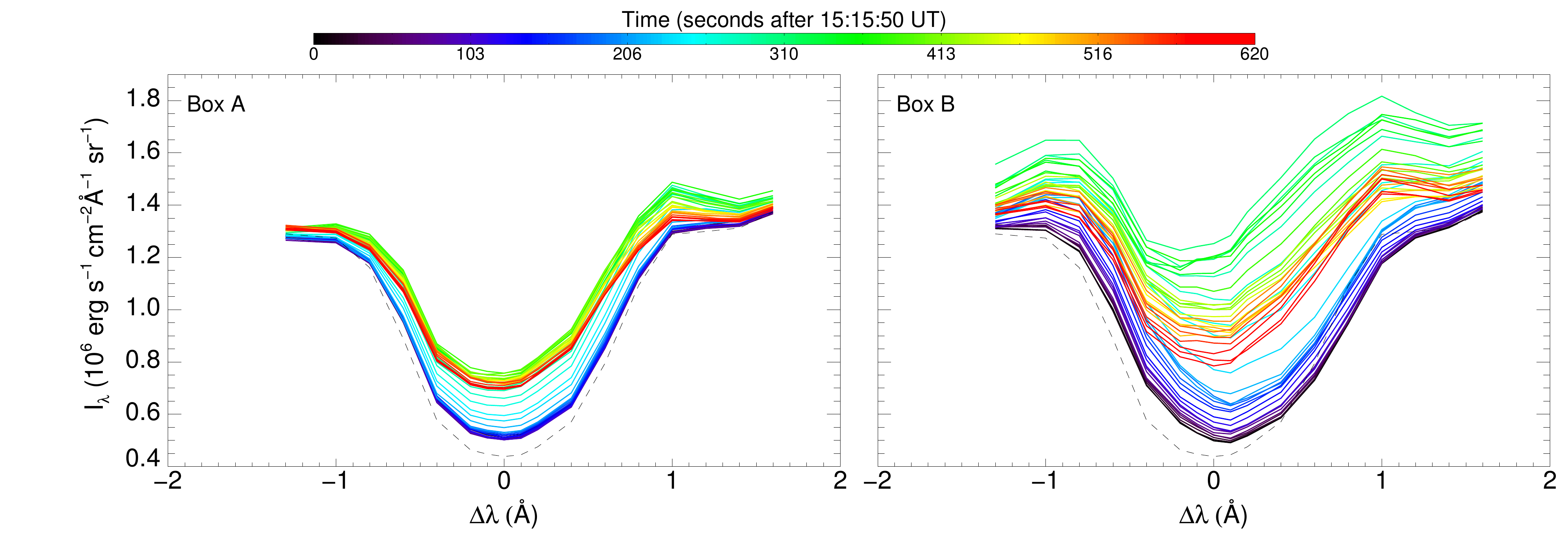}
\caption{Temporal evolution of the observed H$\alpha$ profile in box A (left) and box B (right). The intensities are reported in absolute units. Different colours indicate different times (from black, through blue and green to dark red). The dashed profile results from the average intensity in each box at the beginning of the observing sequence.} 
\label{haprof}
\end{figure*}

Figure \ref{haprof} (\textit{left} panel) displays that in box A the H$\alpha$ line shows a stronger increase in the red wing, compared to the blue wing, and that the line core is shifted blueward. In box B, the line exhibits both red and blue wing enhancements (see Figure \ref{haprof}, \textit{right} panel). The core is more enhanced than in box A but it does not go into emission. In box B the core shows a very small red-shift during the rise phase (up to 200 seconds, see Figure \ref{haprof}, \textit{right} panel), shifting to shorter wavelengths afterwards.
In both boxes A and B, the line does not show a central reversal. \citealt{Deng} reports similar observations, in contrast to typical observations of the H$\alpha$ line profile \citep{Canfield90,delab,KuridzeMathioudakisSimoes:2015}.

\section{RADYN Simulations}

\begin{figure}[ht] 
\centering
\includegraphics[width=0.47\textwidth]{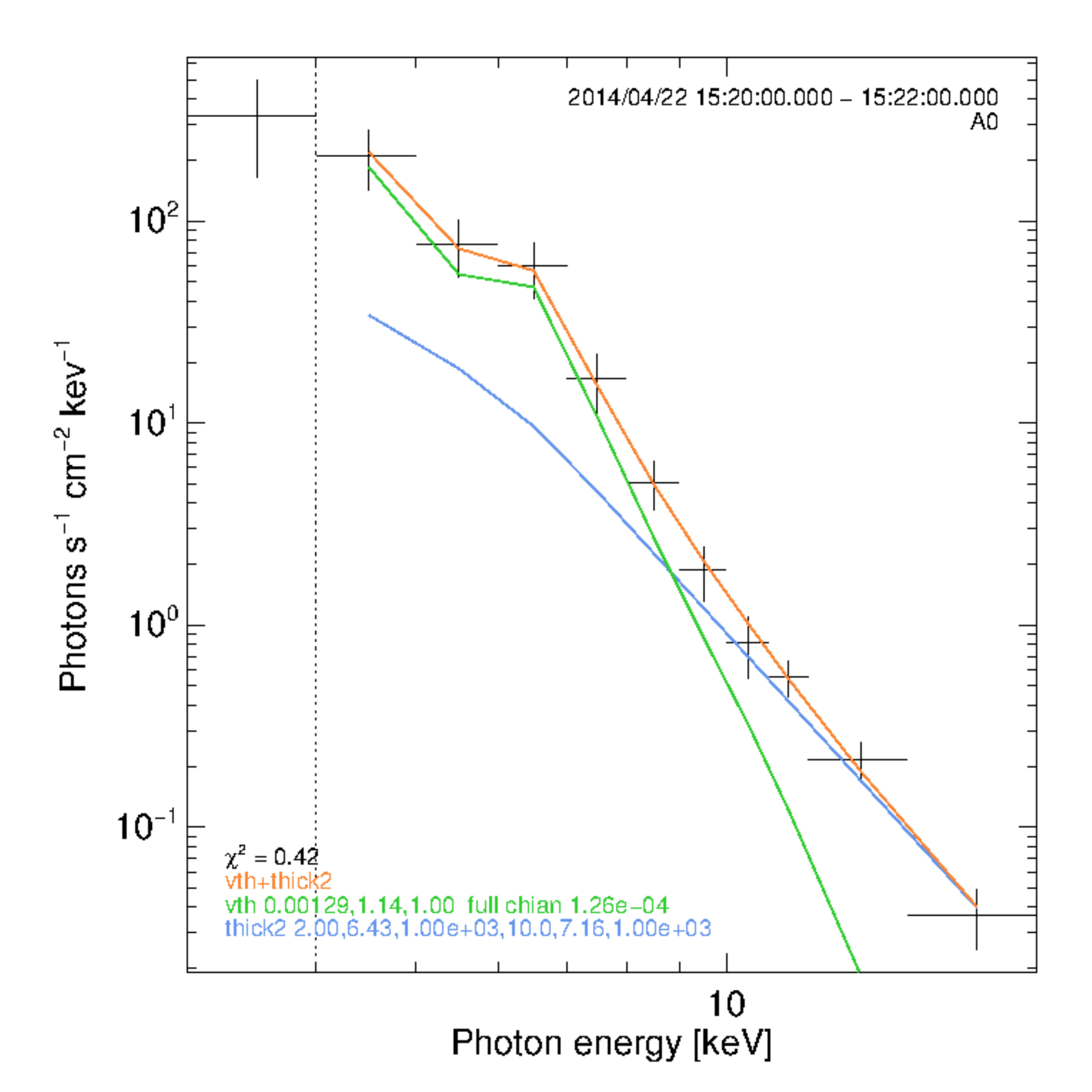}
\caption{RHESSI photon spectra for the HXR over the ROI indicated by purple dashed circle in the Figure \ref{ab1} and integrated for the interval 15:20:00 -- 15:22:00 UT. The orange line is the isothermal plus thick-target model. The isothermal model is the green line, defined by two parameters: emission measure EM and temperature $T$ (see text). The non-thermal thick-target model is the blue line, defined by three parameters: low energy cut-off E$_c$, spectral index $\delta$ and total number of electrons per second above E$_c$.} 
\label{rhessi_spec}
\end{figure}

We used the radiative hydrodynamic code RADYN to compute the H$\alpha$ and H$\beta$ line profiles and to calculate the intensity ratio of the line cores for comparison with the observations. 
Our idea is to adopt a model closest to observations features, whereby we used the RHESSI imaging spectroscopy integrating over 2 minutes during the main impulsive phase (15:20:00 -- 15:22:00 UT). Figure \ref{rhessi_spec} shows RHESSI photon spectra for the HXR source spatially integrated over the region of interest (ROI) displayed in Figure \ref{ab1}. The spectra were fitted with an isothermal plus thick-target model, shown in Figure \ref{rhessi_spec} with orange line, and the resulting parameters are: emission measure EM=1.3$\times$10$^{46}$ cm$^{-3}$; plasma temperature $T$=13 MK; number of electrons per second injected F=2$\times$10$^{35}$ electrons s$^{-1}$ above E=E$_c$=7 keV; spectral index $\delta$=6.4; total non-thermal power P$_{nth}$=2.7$\times$10$^{27}$ erg s$^{-1}$ and the total non-thermal energy E$^{tot}_{nth}$=3$\times$10$^{29}$ erg. From these parameters, two RADYN simulations were employed here, the first one, marked by F9.5, with a peak of energy flux of F$^{max}_{F9.5}$ = 3$\times$$10^{9}$ erg cm$^{-2}$ s$^{-1}$ and a total amount of energy F$^{tot}_{F9.5}$ = 0.3$\times$$10^{11}$ erg cm$^{-2}$; the second one, indicated by F10, with a peak of F$^{max}_{F10}$ = $10^{10}$ erg cm$^{-2}$ s$^{-1}$ and a total amount of energy  F$^{tot}_{F10}$ = $10^{11}$ erg cm$^{-2}$. For both the runs a beam with an isotropic pitch angle distribution in the forward hemisphere was used with the Fokker-Planck solution to the non-thermal electron distribution \citep{AllredKowalskiCarlsson:2015}. A triangular pulse shape heating flux was applied for 20 seconds (with peak after 10 s), and the atmosphere was allowed to relax for 15 additional seconds. The initial atmosphere before the switch-on of the beam was the VAL3C semi-empirical atmosphere \citep{Vernazza}. The electron beam energy distribution is defined as a power-law with a spectral index of $\delta$ = 6 and a low energy cut-off E$_{c}$ = 10 keV. Considering the location of the flare on the solar disk, to simulate the projection effect we used $\mu$=0.23 in both F9.5 and F10 runs.

\begin{figure}[htb] 
\centering
\includegraphics[width=0.5\textwidth]{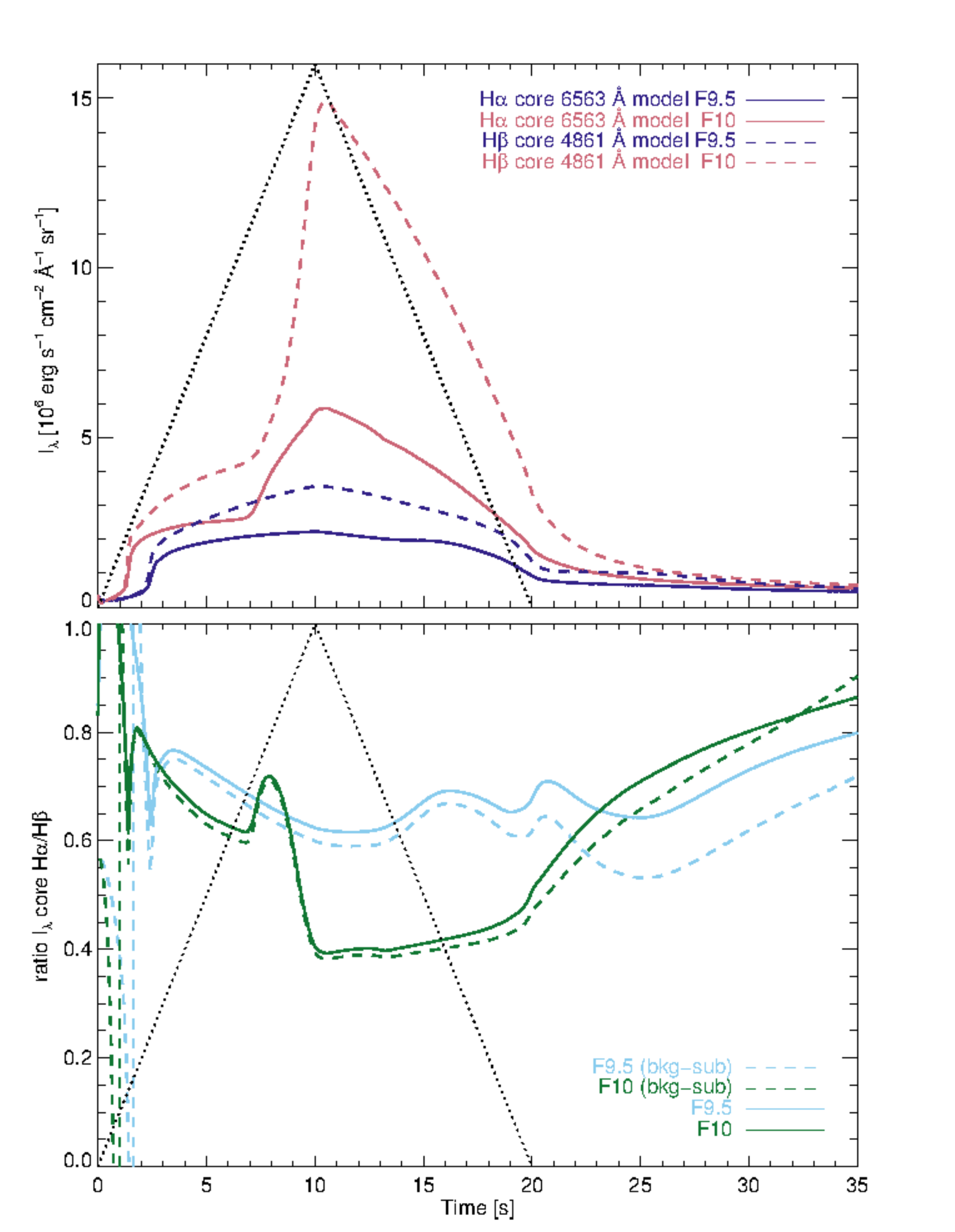}
\caption{Time evolution of the intensity and the ratio of the core in the H$\alpha$ and H$\beta$ lines, for the F9.5 and F10 models (see legend). The black dotted lines show the shape of the energy input.}
\label{ratiohahb1}
\end{figure}

Using the H$\alpha$ and H$\beta$ lines profiles calculated with RADYN, we computed the ratio of the line core intensities, $\lambda=6563$ \AA~ for H$\alpha$ and $\lambda=4861$ \AA~ for H$\beta$, for both models F9.5 and F10, the results of which are shown in Figure \ref{ratiohahb1}. In both models, for the duration of the energy input, the ratio of H$\alpha$ to H$\beta$ is smaller than 1. In F10 the ratio starts around 0.6--0.7 then settles at 0.4 after 10 s, very similar to observed ratios of boxes B and C, while in F9.5 the ratio swings between 0.6--0.8, closer to the observed ratios of box A which is the weakest flare kernel.

\begin{figure}[ht] 
\centering
\includegraphics[width=0.5\textwidth]{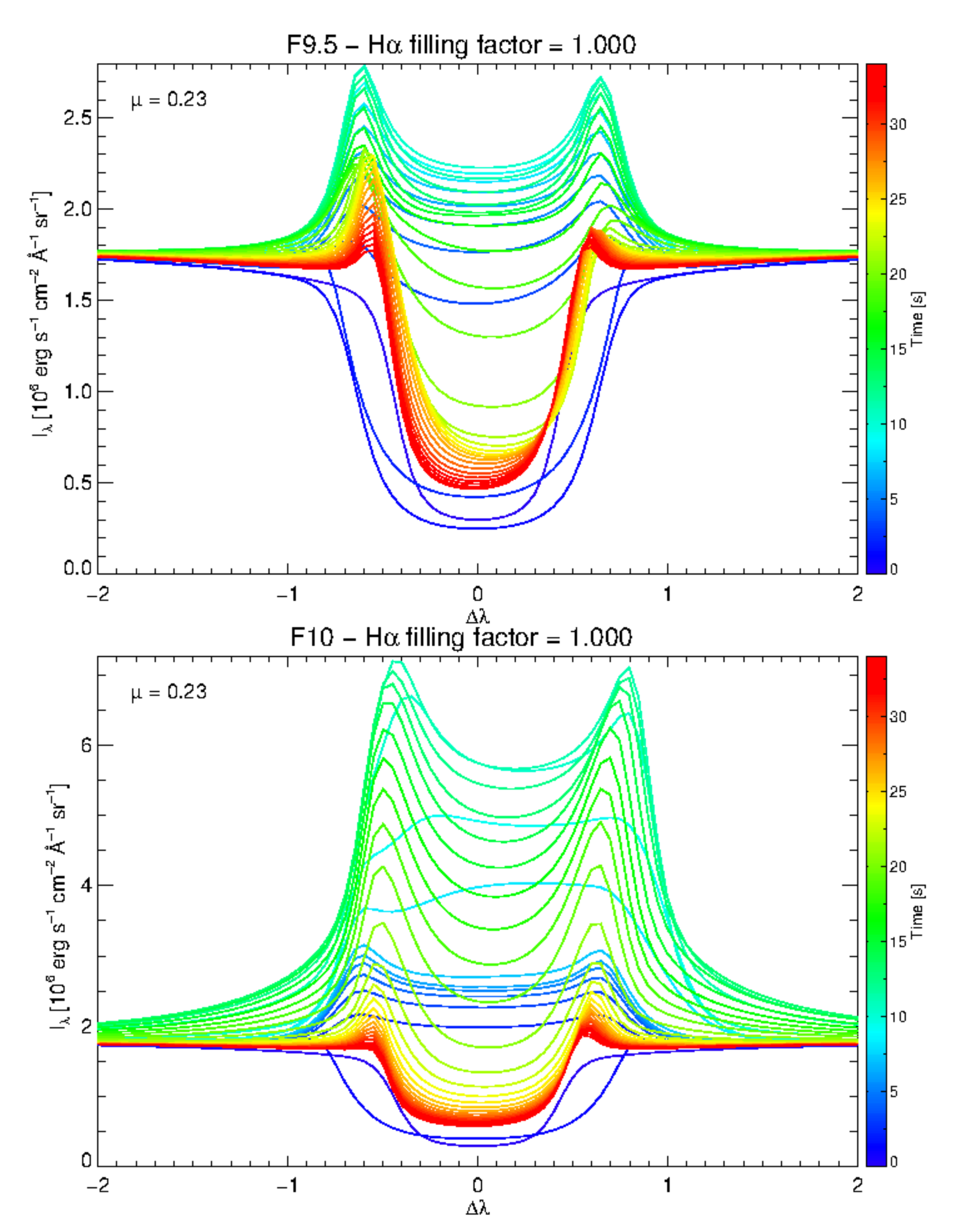}
\caption{Time evolution of the H$\alpha$ line profile calculated with RADYN, for the F9.5 (top) and F10 (bottom) flare models.} 
\label{simhahb1}
\end{figure}

Figure \ref{simhahb1} shows the temporal evolution of the synthesised H$\alpha$ line profile for both F10 and F9.5 flare models. In the F9.5 model, the line, initially in absorption, goes into emission with a clear presence of a central reversal. Through the duration of the energy input, the line wings around $\Delta \lambda \approx -0.5$ are much more pronounced than the line core. During the first 15 seconds of the energy input, the entire line profile is shifted towards longer wavelengths, with the blue wing being stronger than the red wing. From then until the end of the energy input at $t=20$ s, the line shifts slightly redward from the rest wavelength and the wings are more symmetric. When the energy input stops, the intensity of the wings and core decrease substantially, a red-wing asymmetry appears and the core is shifted $\Delta \lambda > 0$.

In the F10 model, the line also goes into emission with a central reversal, with a much more pronounced intensity of the line core with respect to the wing intensities, compared to the F9.5 model. As noted by \cite{KuridzeMathioudakisSimoes:2015}, a red wing asymmetry (i.e., the red wing intensity stronger than the blue wing) develops in the first 5 seconds, before reverting to a blue wing asymmetry after that time. The line core shifts to opposite directions: blueward and then redward. The wing asymmetry is created by an excess of absorption by the moving plasma above the height of formation of the H$\alpha$ line, as pointed out by \cite{KuridzeMathioudakisSimoes:2015}, i.e., a red asymmetry does not necessarily indicate the presence of downward moving plasma, nor does a blue asymmetry indicate upwardly moving plasma. When the energy input ends, the line intensity drops rapidly, the wings become much less pronounced (even disappearing) but the line remains in emission until the end of the simulation.

The synthetic H$\alpha$ line profiles obtained from the RADYN simulations are different from the profiles observed with IBIS, as described earlier. The line profiles in both boxes A and B remain mostly in absorption throughout the event, with a stronger increase in the core intensity than the wings intensities. 

\begin{figure}[ht] 
\centering
\includegraphics[width=0.5\textwidth]{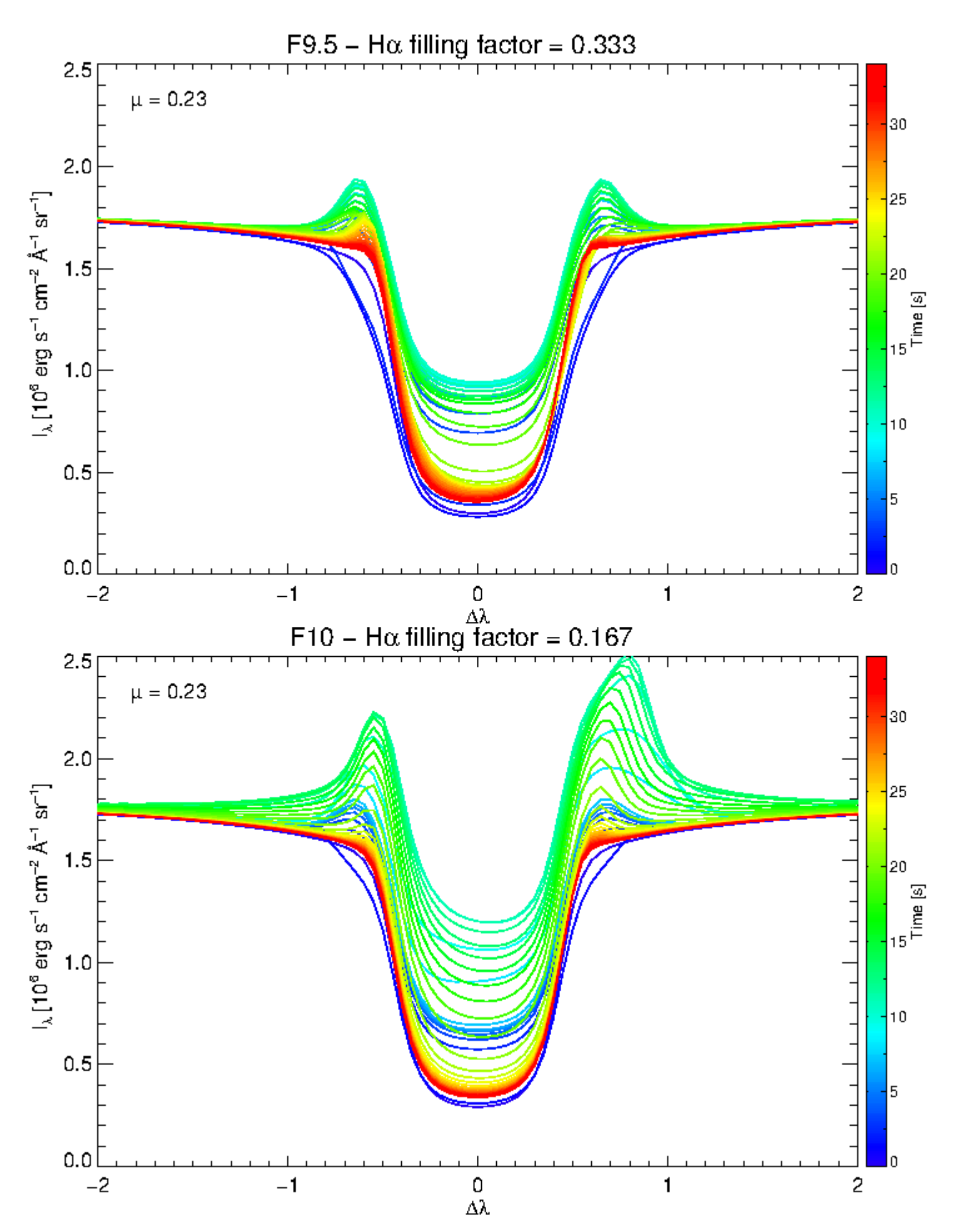}
\caption{Same as Figure \ref{simhahb1}, but simulating a filling factor \textit{ff} smaller than 1 (see text). Top: F9.5 model, \textit{ff}=0.167. Bottom: F10 model, \textit{ff}=0.048.} 
\label{simhahb2}
\end{figure}

It is possible that not all the pixels inside the box are activated by the flares, so the ''filling factor'', defined as \textit{ff}=1/$(N+1)$, is smaller than 1. We have tried to simulate this filling factor effect with the RADYN lines by averaging $N$ times the pre-flare line profile with the flaring lines, namely $I_{\textit{ff}} = \textit{ff}  (I_\mathrm{flare} + NI_\mathrm{pre-flare})$. In practical terms, this brings the calculated lines closer to the observed line shape. For both F9.5 and F10 models, we found solutions that qualitatively reproduce the observations: $N_\mathrm{F9.5}$=2 and $N_\mathrm{F10}$=5 with the filling factor values \textit{ff}$_{F9.5}$=0.333 and ff$_{F10}$=0.167. Figure \ref{simhahb2} shows the line profiles using that filling factor. The plots display that the line is in absorption, prior to a small enhancement of the wings around $\Delta \lambda \pm$1.0 \AA. In F10 there is a small enhancement at $\Delta \lambda \pm $1.0 \AA, and in F9.5 the blue-shift of the line core follows closer the observations than F10. In absolute values of $I_\lambda$, F10 gives a better agreement in $I(\Delta \lambda=0)$, with F9.5 being too weak.

Note that with H$\beta$ we do not have line scans as we do with H$\alpha$. The H$\beta$ images are from the core of the line, and have a relatively broad filter width in comparison to the H$\alpha$ scans. Therefore, RADYN simulation can not be used for a direct comparison for the H$\beta$ images.

We used the F9.5 and F10 to qualitatively explore a weak and a strong case. To determine which of the two runs is closer to observations we have to derive the total amount of energy injected in the two models. RADYN solves equations along one dimension, but the observation results are related to boxes of two dimensions. Our idea is to translate the filling factor discussion in an area value in order to obtain a size for the flaring elements. Comparing the observed and synthetic H$\alpha$ lines profile (see Figure \ref{haprof} and \ref{ratiohahb1}) we obtained a different value of filling factor from F9.5 and F10 runs. Knowing the size of box used in the observations ($5.5''\times5.5''$), we can easily convert them to a simulated flaring area: $\approx 3.2''\times3.2''$ for F9.5 and $\approx 2.2''\times2.2''$ for F10. To estimate the projection effect, from the location of the flare source on the solar disk we obtained $\mu$=0.3 and, using this value, the effective flaring areas are then A$^{eff}_{F9.5}$=17.5$\times$10$^{16}$ cm$^2$ and A$^{eff}_{F10}$=8.8$\times$10$^{16}$ cm$^2$.
From the RHESSI photon spectra fit parameters and using the effecive flaring areas we found F=P$_{nth}$/A$^{eff}_{F9.5}$$\approx$0.15$\times$10$^{11}$ erg s$^{-1}$ cm$^{-2}$ and F=P$_{nth}$/A$^{eff}_{F10}$$\approx$0.31$\times$10$^{11}$ erg s$^{-1}$ cm$^{-2}$. The real problem here is that the RHESSI values are derived from the whole flare (see the ROI in Figure \ref{ab1}), not just one source (box), but anyway the F9.5 model appears closer to the observations.

\section{Conclusions}

The aim of this work was to clarify some aspects related to energy release and redistribution in the chromospheric layer of the solar atmosphere during a solar flare. Therefore, we investigated the chromospheric response to the sudden energy input, locating the sources, the sizes and the eventual offset between flare sources in different wavelengths. Our approach was to look at the features of two chromospheric lines during a C3.3 solar flare, using high resolution ground-based data  acquired during an observing campaign carried out at Dunn Solar Telescope. Although the location of the flare on disk was not ideal and the intensity of the event was relatively low, the uniqueness of this dataset in terms of the resolution (both spatial and temporal), the lines used (both H$\alpha$ scans and H$\beta$ core images) and the fact that all phases of the flare were observed, provides novel insights into the behaviour of the cromosphere during a flare.

In a previous work, \cite{kaspa09} simulated Balmer lines during impulsive flare heating and investigated the correlation between H$\alpha$ and H$\beta$ lines. The authors tried to use H$\alpha$/H$\beta$ ratio to check whether they are sensitive to electron beam presence, i.e., whether they are significantly different if the non-thermal collisional rates are included in the simulations.
In this paper we displayed the lightcurves from observed H$\alpha$ and H$\beta$ lines, where, in the flaring region, the H$\beta$ intensity excess is greater than those of H$\alpha$ during the flare energy input, as noted by \cite{kaspa09}. 

We employed the radiative hydrodynamic code RADYN to compute the synthetic H$\alpha$ and H$\beta$ line profiles to compare them with our observations. For both F9.5 and F10 runs the H$\beta$ intensity is greater than H$\alpha$, while the intensity ratio is around 0.4 in the F10 model after the energy peak, which is in agreement with the observed ratios for the different regions of the flare. Similar ratios from different inputs implies that the H$\alpha$ and H$\beta$ lines are affected similarly by the amount of energy and this indicates that this ratio is sensitive to the amount of energy deposited in the chromosphere.

Our results for the line ratios are in agreement with the findings of \citet{kaspa09}. In \cite{kaspa09}, Flarix gives intensities in erg s$^{-1}$ cm$^{-2}$ Hz$^{-1}$ sr$^{-1}$ and also line ratios are computed from those values, while in this paper we used erg s$^{-1}$ cm$^{-2}$ \AA$^{-1}$ sr$^{-1}$ as $I_\lambda$ units. As confirmed by private communication with the authors, the ratio should be around 0.5 after adjusting to the same units of $I_\lambda$ used in this paper, very close to our results.

Furthermore, by comparing the observed and synthetic H$\alpha$ line profile evolutions, there is good agreement using a ''filling factor'' approach. The simulated H$\alpha$ profiles present a clear central reversal, while the observed line profiles were enhanced during the flare, but remained mostly in absorption. We have interpreted the weakly-enhanced H$\alpha$ line profiles as an effect of a low filling-factor, estimated to be \textit{ff} $\approx$ 0.33 or \textit{ff} $\approx$ 0.17, using models F9.5 and F10, respectively. The simulated intensity ratios, with the pre-flare level subtracted, for the low filling factor cases yield values in the range 0.4--0.5. This is simply because the same filling factor was applied for both H$\alpha$ and H$\beta$ line profiles. Furthermore, converting the filling factor in terms of flaring area we obtained values of total energies for F9.5 and F10 models, indicating the F9.5 closer to the observations.

\begin{figure}[ht] 
\centering
\includegraphics[width=0.22\textwidth]{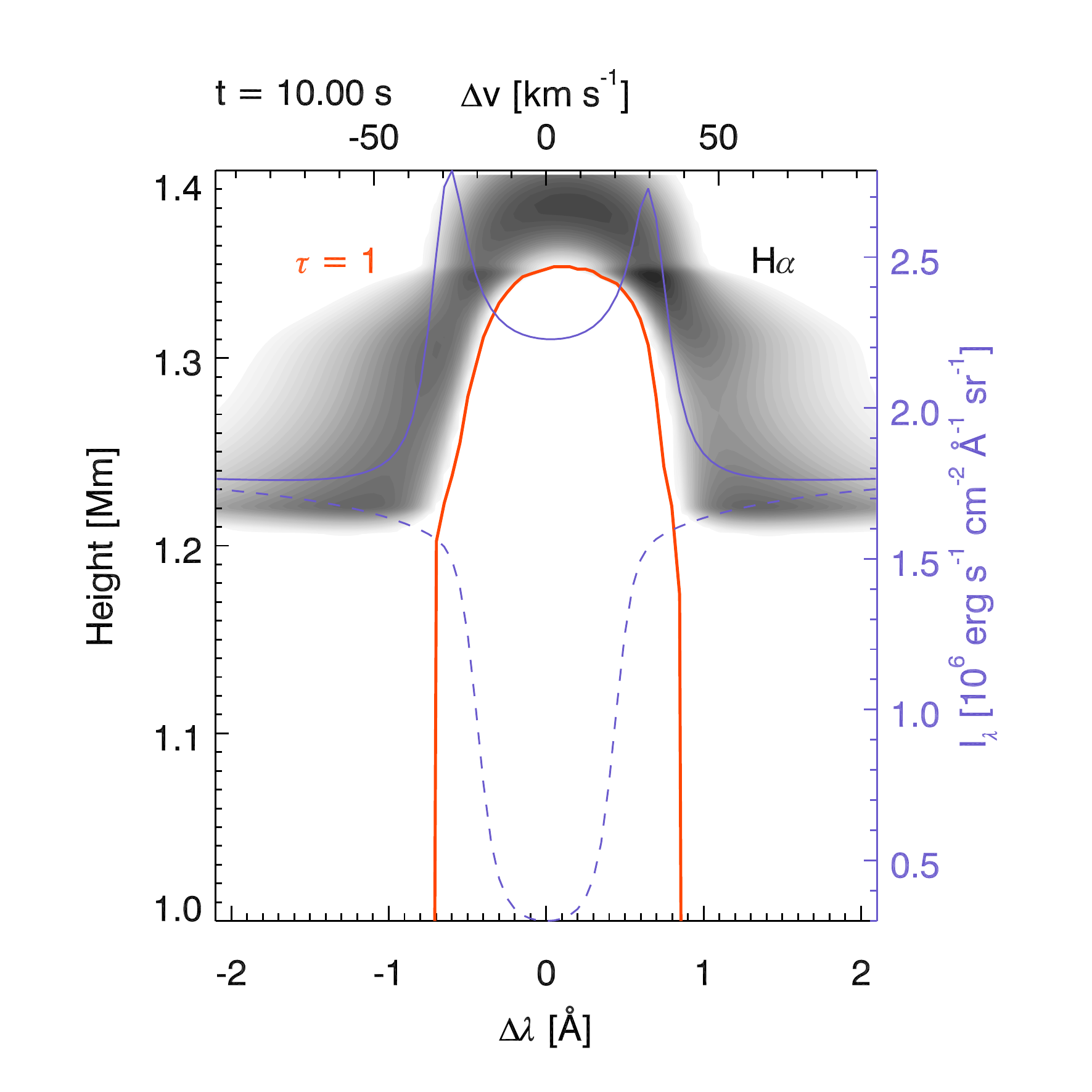}
\includegraphics[width=0.22\textwidth]{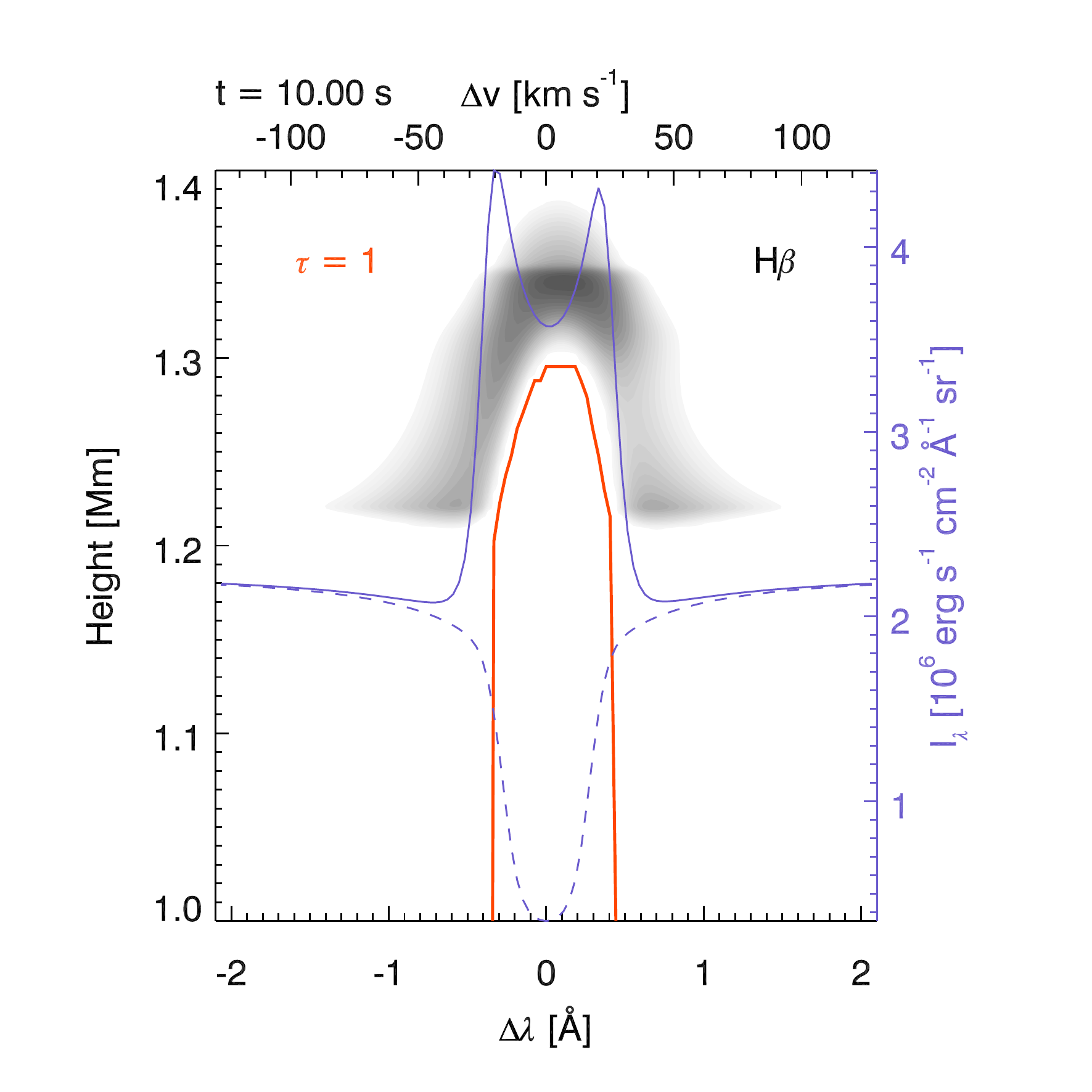}
\caption{Intensity contribution functions for the H$\alpha$ (left) and H$\beta$ (right) lines after 10 s of F9.5 flare heating. The diagrams are plotted in inverse grayscale so that darker shades indicate higher intensities. The line profile is overplotted as a blue line. Red lines indicate the height at which $\tau$=1. Positive velocity corresponds to plasma upflows.} 
\label{con_f}
\end{figure}

Concerning the analysis related to the spatial offset, because the flare is located close to the limb, we can read the results as diagnostic of the formation heights of the two line-cores. Figure \ref{con_f} displays the line contribution functions \citep{carlsson97} after 10 s of F9.5 flare input, where the H$\alpha$ core formation height is higher of about 40 km with respect to the H$\beta$ core (wings formation heights are the same). The observed spatial separation is in qualitative agreement with the RADYN simulation, the observed spatial offset is much larger but this may be due to one dimensional limit of RADYN code, so it's hard to make an actual comparison with imaging observations. Therefore, a possible suggestion for future observations in both spectral lines would be to search for flaring active regions very close to the solar limb, in order to further investigate the spatial offset.

\acknowledgments

The research leading to these results has received funding from the European Community's Seventh Framework Programme (FP7/2007-2013) under grant agreement no. 606862 (F-CHROMA). This research has received funding from the European Commission’s Seventh Framework Programme under the grant agreement no. 312495 (SOLARNET project). The IBIS and ROSA data analyzed in this paper were acquired in the framework of SOLARNET service mode. This work was also supported by the Italian MIUR-PRIN grant 2012P2HRCR on The active Sun and its effects on space and Earth climate, by Space Weather Italian COmmunity (SWICO) Research Program, by the Istituto Nazionale di Astrofisica (PRIN INAF 2010/2014), and by the Universit\'a degli Studi di Catania.

PJAS and LF acknowledge support from grant ST/L000741/1 made by the UK's Science and Technology Facilities Council.

This research was supported by the Research Council of Norway through the grant ''Solar Atmospheric Modelling'' and through grants of computing time from the Programme for Supercomputing.

The RADYN models used in this work are part of the F-CHROMA database of solar flare models, available at \url{www.fchroma.org}, under "data access". We would also like to thank Jana K\v{a}sparov\'{a} for discussing their 2009 results in detail.

\end{document}